\documentclass{article}
\usepackage[a4paper, total={6in, 8in}]{geometry}
\usepackage[utf8]{inputenc}
\usepackage{setspace}
\RequirePackage{amsthm,amsmath,amsfonts,amssymb,bm}
\RequirePackage[authoryear]{natbib}
\RequirePackage[colorlinks,citecolor=blue,urlcolor=blue]{hyperref}
\RequirePackage{graphicx}
\usepackage{mathtools}
\usepackage{authblk}

\numberwithin{equation}{section}
\theoremstyle{plain}

\theoremstyle{remark}
\newtheorem{rmk}{Remark}

\providecommand{\keywords}[1]
{
  \small	
  \textbf{\textit{Keywords---}} #1
}

\title{A population-aware retrospective regression to detect genome-wide variants with sex difference in allele frequency}

\author[1]{Zhong Wang}
\author[2,3,4]{Andrew D. Paterson}
\author[3,5]{Lei Sun}

\affil[1]{Department of Statistics and Data Science, Faculty of Science, National University of Singapore, Singapore}

\affil[2]{Genetics and Genome Biology, The Hospital for Sick Children, Toronto, Ontario, Canada}

\affil[3]{Biostatistics Division, Dalla Lana School of Public Health, University of Toronto, Toronto, Ontario, Canada}

\affil[4]{Epidemiology Division, Dalla Lana School of Public Health, University of Toronto, Toronto, Ontario, Canada}

\affil[5]{Department of Statistical Sciences, Faculty of Arts and Science, University of Toronto, Toronto, Ontario, Canada}

\begin{document}
\maketitle


\begin{abstract}
Sex difference in allele frequency is an emerging topic that is critical to our understanding of ascertainment bias, as well as data quality particularly of the largely overlooked X chromosome. To detect sex difference in allele frequency for both X chromosomal and autosomal variants, existing methods are conservative when applied to samples from multiple ancestral populations, such as African and European populations. Additionally, it remains unexplored whether the sex difference in allele frequency differs between populations, which is important to trans-ancestral genetic studies. We thus developed a novel retrospective regression-based testing framework to provide interpretable and easy-to-implement solutions to answer these questions. We then applied the proposed methods to the high-coverage whole genome sequence data of the 1000 Genomes Project, robustly analyzing all samples available from the five super-populations. We had 76 novel findings by recognizing and modeling ancestral differences. 

\end{abstract}


\section{Introduction} \label{sec:1}
The omnipresent genome-wide association studies (GWAS) routinely exclude the important X chromosome, despite the availability of the data \citep{wise_exclusion_2013}. To overcome this major research gap, previous work focused on developing powerful association tests tailored for the X chromosomal variants \citep{chen_x_2021}. However, the earlier work implicitly assumed that the standard data quality control procedure, including for example missing rate, cryptic relatedness and Hardy--Weinberg equilibrium (HWE) \citep{anderson2010data,marees_tutorial_2018}, would lead to good quality data, forgetting that most of the upstream genotype calling bioinformatic tools and imputation methods are also autosome-centric, even if the X chromosome is included in the analysis \citep{konig2014include,das2016next,taliun2021sequencing,browning2021fast}. 

The sex difference in minor allele frequency (sdMAF) of single nucleotide polymorphism (SNP) is an important, but previously neglected indicator of data abnormality in GWAS \citep{wang_major_2022}. Recently, among the SNPs presumed to be of high quality in the phase 3 data \citep{the_1000_genomes_project_consortium_global_2015} and high-coverage whole genome sequence data \citep{byrska-bishop_high-coverage_2022} of the 1000 Genomes Project, many with genome-wide significant sdMAF were identified \citep{wang_major_2022}. In particular, the proportion of SNPs with significant sdMAF was much higher on the X chromosome than autosomes. 

Despite the recent success at detecting both autosomal and X chromosomal SNPs with significant sdMAF, the existing test is known to be conservative when it is applied to datasets that include multiple populations. In the meantime, understanding population heterogeneity is an increasingly important topic, as we collect more diverse and inclusive data. For example, the 1000 Genomes Project \citep{the_1000_genomes_project_consortium_global_2015,byrska-bishop_high-coverage_2022} includes 2,504 individuals from five super-populations, including African (AFR), American (AMR), East Asian (EAS), European (EUR), and South Asian (SAS). Previous work \citep{wang_major_2022} analyzed all 2,504 individuals jointly without adjusting for population effect, or equivalently used the standard meta-analysis \citep{willer2010metal} to combine the sdMAF summary statistics across the five populations. Although the existing test is valid in the presence of the known population difference in MAF \citep{byrska-bishop_high-coverage_2022}, as ``sdMAF test detects the difference in MAF between females and males, not the difference in MAF between populations" \citep{wang_major_2022}, it is not powerful if the {\it direction} of sdMAF differs between populations. Additionally, the existing test was not designed to detect SNPs with significant {\it difference in sdMAF between populations}, which is also of interest to the genetic research community.   


In this paper, we propose a population-aware, retrospective regression-based testing framework that i) includes the existing sdMAF test as a special case when analyzing a single population, ii) performs a new multi-population sdMAF test that is robust to sdMAF heterogeneities (in magnitude and direction) between populations, iii) conducts a new between-population sdMAF test, comparing the extent of difference in sdMAF between populations, iv) is flexible enough to adjust for covariate effects (e.g.\ age and subtler population differences), and finally v) is computational scalable to analyze large-scale data in terms of both sample size and the number of variants analyzed. 


In Section \ref{sec:2}, we introduce the statistical framework and the related multi-population and between-population sdMAF tests  for both X-chromosomal and autosomal SNPs.  In Section \ref{sec:3}, we evaluate the accuracy of proposed methods via realistically simulated data, generated based on the high-coverage data of the 1000 Genomes Project. In Session \ref{sec:4}, we apply the proposed tests to the high-coverage data of the 1000 Genomes Project, and we report new multi-population sdMAF findings missed by the earlier conservative test, as well as novel between-population sdMAF findings. In Session \ref{sec:5}, we discuss implications of our findings, limitations of the proposed methods, as well as some promising future extensions. {Finally, in Section \ref{sec:6}, we provide the links to the implemented program, the codes used for our simulation studies, as well as the specific version of the publicly available 1000 Genomes Project data used for our application study.}

\section{The retrospective regression-based sdMAF testing framework} \label{sec:2}

\subsection{Notation}
\label{subsection:notation} 

Consider a bi-allelic SNP, the most common type of genetic variations, with two alleles denoted as $b$ and $B$, where  $B$ is the minor allele with population allele frequency $p\leq 0.5$ by convention. Also by convention, for a SNP from one of the 22 autosomes, the genotypes $bb,Bb,BB$ are coded $0$, $1$ and $2$, for both females and males (Table \ref{tab1}). 

\begin{table}[h!]
\begin{center}
\begin{tabular}{c  ccc} \hline 
Genotypes & Autosomes & Xchr PAR & Xchr NPR \\ \hline
& Female and Male & Female and Male & Female \\ \hline
$bb$ & 0 & 0 & 0\\ 
$Bb$ & 1 & 1 & 1\\
$BB$ & 2 & 2 & 2\\ \hline
 & & & Male \\ \hline
$b$ & NA & NA & 0 \\ 
$B$ & NA & NA & 2 \\  \hline
\end{tabular}
\end{center}
\caption{{\bf Genotype coding for a bi-allelic SNP, stratified by genomic region and sex.}  PAR is the pseudoautosomal region of the X chromosome (Xchr), and NPR is the non-pseudoautosomal regions of the Xchr. For a NPR SNP, male's genotypes are hemizygous, and the corresponding genotypes $b$ and $B$ are coded as $0$ and $2$ by convention.} 
\label{tab1}
\end{table}

The same genotype coding is used for a SNP from the  pseudoautosomal regions (PAR) of the X chromosome, which recombines with the Y chromosome in males and the other X chromosome in females.  In contrast, for a SNP from the non-pseudoautosomal regions (NPR) of the X chromosome, the genotypes $bb,Bb,BB$ in females are coded $0$, $1$ and $2$, and the (hemizygous) genotypes $b$ and $B$ in males are coded $0$ and $2$, by convention \citep{purcell2007plink}. We note that the $0$ and $2$ coding assumes X-inactivation, an important assumption for studying genotype-phenotype relationship \citep{chen_x_2021}. However, for the purpose of this study, the choice of the coding is important for a different reason, which will become clear in Section \ref{subsection:model}.   

Let $G$ be the genotype coding, as described above, for each individual from a random sample of size $n$, $Sex$ be the sex indicator  ($Sex=0$ for a male 
 and $Sex=1$ for a female without loss of generality), and $POP_k$ be the population indicator for population $k$, where $k=1,2,\dots,K-1$. To ease notation we use $k=0$ to represent the baseline population, which means $POP_k=0, \forall k=1,2,\dots,K-1$. Additionally, we use subscripts $_f$ and $_m$ to represent females and males respectively, and double subscripts $_{f,k}$ and $_{m,k}$ to represent, respectively, females in population $k$ and males in population $k$. 

Finally, let $\delta$ be the Hardy-Weinberg disequilibrium (HWD) parameter, and $p$ be the allele frequency of allele $B$. Then, $\delta= \text{freq}(BB) - p^2$, the difference between the observed genotype frequency and expected under HWE \citep{crow1970introduction,zhang_generalized_2022}. Note that, after some simple algebra, the HWD $\delta$ can be equivalently defined as $\delta= \text{freq}(bb) - (1-p)^2$ or $\delta= (2p(1-p)-\text{freq}(Bb))/2$. Worth noting here is also the fact that, for the X chromosome, the concept of HWD does not apply to {\it a NPR SNP in males}, as the two genotypes ($b$ and $B$) are hemizygous, and the corresponding genotype frequencies are the allele frequencies. 

\subsection{The general sdMAF testing framework}
\label{subsection:model} 

To test for multi-population sdMAF or between-popuation sdMAF, we propose the following retrospective regression framework,
\begin{equation}\label{mult_model}
    G=\alpha+\gamma*Sex+\sum_{k=1}^{K-1}\gamma_k*POP_k+\sum_{k=1}^{K-1}\eta_k*Sex*POP_k+\dots+\epsilon_{Sex, POP_k},
\end{equation}
where $\epsilon_{Sex, POP_k}\sim N(0,\sigma_{f,k}^2)$ for females and $N(0,\sigma_{m,k}^2)$ for males in population $k$. It is crucial that $\epsilon_{Sex, POP_k}$ is sex- and population-specific, because variance of a genotype depends on MAF, while MAF is hypothesized here to differ between sexes and also known to differ between populations \citep{byrska-bishop_high-coverage_2022}. The $\mathord{\cdot}\mathord{\cdot}\mathord{\cdot}$ in the model stands for other covariates (e.g.\ age and smoking status) if available, but they are omitted for now to ease notation but without loss of generality. 

The use of the Gaussian model may appear to be counter-intuitive, but its use for an discrete outcome is well justified by the earlier theoretical work of \cite{chen_score_1983} and the more recent work in genetic association studies \citep{zhang_generalized_2022,https://doi.org/10.1002/cjs.11729} and causal inference via Mendelian randomization \citep{ye2021genius}.  Additionally, the Gaussian model is easy-to-interpret, as $E(G)=2p$, where $p$ is the MAF, using the conventional genotype coding as shown in Table \ref{tab1}. 
Thus, the proposed regression examines how MAF depends on sex (i.e.\ single-population sdMAF), population (i.e.\ multi-population sdMAF), and their interacting effect (i.e.\ between-population sdMAF). Finally, we show below that the proposed testing framework includes the existing sdMAF test as a special case.

\subsection{Single-population sdMAF analysis} \label{subsection:2.1} 

If all individuals come from a single ancestry group, the regression model is simplified to
\begin{equation}\label{sg_model}
    G=\alpha+\gamma*Sex+\epsilon_{Sex},
\end{equation}
where $\epsilon_{Sex} \sim N(0,\sigma_{f}^2)$ for females and $N(0,\sigma_{m}^2)$ for males. Given the genotype coding in Table \ref{tab1}, $E(G)=2p$, {\it regardless of sex or genomic region} (autosomes or PAR and NPR of the X chromosome).  Recall $Sex=0$ for a male and $Sex=1$ for a female. Thus, in model \eqref{sg_model}, the simplified version of \eqref{mult_model}, the regression coefficient parameters represent;
\begin{itemize}
    \item $\alpha$: two times the MAF in males; 
    \item $\gamma$: two times the difference in MAF between females and males, i.e.\ sdMAF.
\end{itemize}

Additionally, when analyzing a SNP from the autosomes or X chromosomal PAR, it is straightforward to obtain the maximum likelihood estimates (MLEs) of the regression parameters,
\begin{equation}\label{sg_mle_A}
    \begin{aligned}
        \hat\alpha & = 2\hat p_m, \\
        \hat\gamma & = 2(\hat p_f - \hat p_m), \\
        \hat\sigma_f^2 & = 2(\hat p_f(1-\hat p_f) + \hat\delta_f), \: \text{ and} \\
        \hat\sigma_m^2 & = 2 (\hat p_m(1-\hat p_m) + \hat\delta_m),
    \end{aligned}
\end{equation}
where the estimated allele frequencies and amount of HWD are given by 
\begin{equation}\label{sg_mle_A2}
  \begin{aligned}
\hat p_f &= \frac{2n_f(BB)+n_f(Bb)}{2n_f}\: \text{ and } \: \hat\delta_f = \frac{n_f(BB)}{n_f} - \hat p_f^2,  \: \text{ and}\\ 
\hat p_m & = \frac{2n_m(BB)+n_m(Bb)}{2n_m}  \: \text{ and } \: \hat\delta_m = \frac{n_m(BB)}{n_m} - \hat p_m^2,
  \end{aligned}
\end{equation}
respectively for females and males.

Importantly, when analyzing a SNP from the X chromosomal {\it NPR}, 
\begin{equation}\label{sg_mle_X}
    \begin{aligned}
        \hat p_m & = \frac{n_m(B)}{n_m}, \: \text{ and} \\
        \hat\sigma_m^2 & = 4\hat p_m(1-\hat p_m),
    \end{aligned}
\end{equation}
as males only have one copy of the X chromosome. Compared to a PAR or an autosomal SNP, the allele frequency estimation clearly differs for a NPR SNP. Additionally, the concept of HWD is ill-defined for a NPR SNP in males. Thus, $\hat\sigma_m^2$ in \eqref{sg_mle_X} does {\it not} include the HWD estimate $\hat \delta_m$. Finally, the number of alleles of an NPR SNP is half of that of a PAR SNP in males, which means the sample size for frequency estimation is smaller. Thus, $\hat\sigma_m^2$ has a factor of 4 in \eqref{sg_mle_X} instead of 2 in \eqref{sg_mle_A}.

To test for sdMAF, we test 
$$H_0:\gamma=0.$$ 
For an autosomal or X chromosomal PAR SNP, the corresponding Wald statistic is,
\begin{equation}\label{sg_auto}
    W_{\text{sdMAF,A}} = \frac{(\hat p_f - \hat p_m)^2}{\frac{1}{2n_f}(\hat p_f(1-\hat p_f)+\hat\delta_f) + \frac{1}{2n_m}(\hat p_m(1-\hat p_m)+\hat\delta_m)} \stackrel{H_0}{\sim} \chi_1^2.
\end{equation}
In contrast, for a X chromosomal NPR SNP, 
\begin{equation}\label{sg_npr}
    W_{\text{sdMAF,X}} = \frac{(\hat p_f - \hat p_m)^2}{\frac{1}{2n_f}(\hat p_f(1-\hat p_f)+\hat\delta_f) + \frac{1}{n_m}\hat p_m(1-\hat p_m)} \stackrel{H_0}{\sim} \chi_1^2.
\end{equation}
The detailed derivations for \eqref{sg_auto} and \eqref{sg_npr} are provided as part of the Supplementary Materials.



\begin{rmk}
The expressions for both \eqref{sg_auto} and \eqref{sg_npr} are identical to the test statistics used in \cite{wang_major_2022}, which were derived intuitively by comparing two sample proportions, $\hat p_f$ and $\hat p_m$. Thus, the proposed regression framework includes the earlier testing method as a special case when analyzing one population.  
\end{rmk}

In the presence of multiple populations, all samples can be analyzed jointly without adjusting for population differences. Unlike the traditional genotype-phenotype association testing \citep{price2006principal}, the subsequent sdMAF test does not have inflated type I error. Instead, it is conservative and at the cost of reduced power, because HWD $\delta$ in the variance is overestimated when ignoring population structure; see Supplementary Note S1 of \cite{wang_major_2022}. To optimally analyze multiple populations, the regression nature of the proposed approach makes method extension both principled and straightforward, which we study next. 

\subsection{Multi-population sdMAF analysis}
Suppose we have a sample consists of individuals from $K$ different ancestry groups, such as the 1000 Genomes Project sample from the five super-populations (AFR, AMR, EAS, EUR, and SAS) \citep{byrska-bishop_high-coverage_2022}. Consider 
\begin{equation}\label{mult_model2}
    G=\alpha+\gamma*Sex+\sum_{k=1}^{K-1}\gamma_k*POP_k+\sum_{k=1}^{K-1}\eta_k*Sex*POP_k+\epsilon_{Sex, POP_k}, 
\end{equation}
where, as noted earlier, $POP_k, k=1,2,\dots,K-1$ are the super-population indicators, and $k=0$ being the baseline super-population. As model \eqref{mult_model2} includes interactions, the regression coefficient parameters are harder to interpret than those in \eqref{sg_model}. Nevertheless, it is not difficult to see that conceptually; 
\begin{itemize}
    \item $\alpha$: two times the MAF in males in the baseline super-population,
    \item $\gamma$: two times the difference in MAF between females and males in the baseline super-population, i.e. sdMAF in the baseline super-population.
    \item $\gamma_k$: two times the difference in MAF in males between super-population $k$ and the baseline super-population, $k=1,\ldots K-1$.
    \item $\eta_k$: two times the difference in sdMAF between super-population $k$ and the baseline super-population, i.e.\ between-super-population sdMAF comparison, $k=1,\ldots K-1$.
\end{itemize}

Define the set of all relevant parameters as $\theta=(\alpha,\gamma,\gamma_1,\eta_1,\dots,\gamma_{K-1},\eta_{K-1},\bm\sigma')'$, where $\bm\sigma=(\sigma_{m,0},\sigma_{f,0},\dots,\sigma_{m,K-1},\sigma_{f,K-1})'$, and $\sigma_{m,0}$ and $\sigma_{f,0}$ represent the variance for the baseline population when $POP_k=0, \forall k=1,2,\dots,K-1$.   The MLE of ${\bm\sigma}$ is independent from the MLEs of the other parameters in $\theta$. Thus, to conduct tests for $\gamma$'s and $\eta$'s, the calculation of the corresponding Wald statistics requires only the Fisher information matrix related to $(\alpha,\gamma,\gamma_1,\eta_1,\dots,\gamma_{K-1},\eta_{K-1})'$. We denote such partial matrix by $I_n^*(\theta)$. 

Note that $I_n^*(\theta)$ is a $2K$ by $2K$ block matrix with non-zero entries in the main diagonal, the first row and the first column. After some algebra, we can obtain the inverse of $I_n^*(\theta)$,  
\begin{equation}
    I_n^*(\theta)^{-1} = \left(\begin{array}{cccc}
                 B_0^{-1}, & -B_0^{-1} & -B_0^{-1}, &  \cdots  \\
                 -B_0^{-1}, &  B_1^{-1}+B_0^{-1}, & B_0^{-1}, &  \cdots  \\
                 -B_0^{-1}, &  B_0^{-1}, & B_2^{-1}+B_0^{-1}, &  \cdots  \\
                 \vdots & \vdots &  \vdots   & \ddots   
              \end{array}\right),
\end{equation}
where 
\begin{equation*}
    B_k^{-1} = \left(\begin{array}{cc}
        \frac{\sigma_{m,k}^2}{n_{m,k}}, & -\frac{\sigma_{m,k}^2}{n_{m,k}} \\
        -\frac{\sigma_{m,k}^2}{n_{m,k}}, & \frac{\sigma_{m,k}^2}{n_{m,k}} + \frac{\sigma_{f,k}^2}{n_{f,k}}
    \end{array}\right).
\end{equation*}
The derivation details are provided in the Supplementary Materials, but we note here that the block feature of  $I_n^*(\theta)^{-1}$ is a direct result of specifying a sex- and population-dependent error model (\ref{mult_model2}).

When analyzing an autosomal or X chromosomal PAR SNP, the global MLEs are,
\begin{equation}\label{mult_mle_A}
    \begin{aligned}
    \hat\alpha & = 2\hat p_{m,0}, \\
    \hat\gamma & = 2 (\hat p_{f,0} - \hat p_{m,0}), \\
    \hat\gamma_k & = 2 (\hat p_{m,k} - \hat p_{m,0}), \\
    \hat\eta_k & = 2(\hat p_{f,k} - \hat p_{m,k}) - 2(\hat p_{f,0} - \hat p_{m,0}), \\
    \hat\sigma_{f,k}^2 & = 2 (\hat p_{f,k}(1-\hat p_{f,k}) + \hat\delta_{f,k}), \: \text{ and}\\
    \hat\sigma_{m,k}^2 & = 2 (\hat p_{m,k}(1-\hat p_{m,k}) + \hat\delta_{m,k}),
    \end{aligned}
\end{equation}
where $k=1,2,\dots,K-1$; for $\hat\sigma^2_{f,k}$ and $\hat\sigma^2_{m,k}$, $k$ includes $k=0$ for the baseline population. The sex- and population-stratified MAF and HWD estimates ($\hat p_{f,k}$, $\hat p_{m,k}$, $\hat\delta_{f,k}$, and $\hat\delta_{m,k}$) follow expressions in (\ref{sg_mle_A2}) for one homogeneous population, with appropriate subscript $_k$ for population $k$, $k=0,1,\ldots, K-1$.

When analyzing a X chromosomal NPR SNP, the expressions of $\hat \alpha$, $\hat \gamma$, $\hat \gamma_{k}$, $\hat \eta_{k}$, and $\hat\sigma_{f,k}^2$ are identical to those above. However, 
\begin{equation}\label{mult_mle_X}
    \begin{aligned}
    \hat\sigma_{m,k}^2 & = 4\hat p_{m,k}(1-\hat p_{m,k}),
    \end{aligned}
\end{equation}
and the male MAF must be estimated appropriately, $\hat p_{m,k} = {n_{m,k}(B)}/{n_{m,k}}$ for population $k$, $k=0,1,\ldots, K-1$.

Given the MLEs of the regression coefficients in (\ref{mult_mle_A}), it is clear that testing for sdMAF across the $K$ populations, while allowing for population-specific MAF and HWD, is to test $H_0:\gamma=\eta_{1}+\gamma=\dots=\eta_{K-1}+\gamma=0$, or equivalently,
$$H_0:\gamma=\eta_{1}=\dots=\eta_{K-1}=0.$$

For an autosomal or X chromosomal PAR SNP, the corresponding Wald test statistic is
\begin{equation}\label{sdMAF_multtest1_A}
    W_{\text{sdMAF,A}}^{\scriptscriptstyle{(0,1,\dots,K-1)}} =\sum_{k=0}^{K-1} W_{\text{sdMAF,A}}^{\scriptscriptstyle{(k)}}\stackrel{H_0}{\sim} \chi_K^2,
\end{equation}
where
\begin{equation}\label{sg_auto2}
    W_{\text{sdMAF,A}}^{\scriptscriptstyle{(k)}} = \frac{(\hat p_{f,k} - \hat p_{m,k})^2}{\frac{1}{2n_{f,k}}(\hat p_{f,k}(1-\hat p_{f,k})+\hat\delta_{f,k}) + \frac{1}{2n_{m,k}}(\hat p_{m,k}(1-\hat p_{m,k})+\hat\delta_{m,k})}.
\end{equation}
For a X chromosomal NPR SNP, the multi-population sdMAF test has the same form as \eqref{sdMAF_multtest1_A}, 
\begin{equation}\label{sdMAF_multtest1_X}
   W_{\text{sdMAF,X}}^{\scriptscriptstyle{(0,1,\dots,K-1)}}= \sum_{k=0}^{K-1} W_{\text{sdMAF,X}}^{\scriptscriptstyle{(k)}}\stackrel{H_0}{\sim} \chi_K^2,
\end{equation}
but each element is replaced by NPR-suitable variance for males,
\begin{equation}\label{sg_npr2}
    W_{\text{sdMAF,X}}^{\scriptscriptstyle{(k)}} = \frac{(\hat p_{f,k} - \hat p_{m,k})^2}{\frac{1}{2n_{f,k}}(\hat p_{f,k}(1-\hat p_{f,k})+\hat\delta_{f,k}) + \frac{1}{n_{m,k}}\hat p_{m,k}(1-\hat p_{m,k})}.
\end{equation}

\begin{rmk}
The proposed multi-population sdMAF test, aggregating sdMAF testing information across $K$ different populations/groups, is fundamentally different from the classical fixed-effect or random-effects meta-analysis \citep{willer2010metal,lin2010relative,borenstein2021introduction}. This is because the proposed multi-population sdMAF tests \eqref{sdMAF_multtest1_A} and \eqref{sdMAF_multtest1_X} are omnibus, robust to opposite directions of sdMAF between the $K$ populations, as the estimated sdMAF, $(\hat p_{f,k} - \hat p_{m,k})$, is squared before the aggregation, akin to quadratically aggregating association information across $K$ rare variants \citep{derkach2014pooled}. Although this is at the cost of reduced power if all directions (and magnitudes) of sdMAF are consistent between the $K$ populations, the reduction is limited even in the worst-case scenario, which we will illustrate in the application to the 1000 Genomes Project data \citep{byrska-bishop_high-coverage_2022} in Section \ref{sec:4}.
\end{rmk}


\subsection{Between-population sdMAF comparison}
The proposed regression framework can naturally lead to tests that compare sdMAF between different populations. We start with pairwise between-population comparison test. Let populations $k$ and $l$ be the two populations of interest, we have two (equivalent) testing strategies. 

One strategy is to (re)define the population indicators, so that say population $k$ is the baseline population. Then we test
$$H_0:\eta_l=0,$$
as $\eta_l$ captures the difference in sdMAF between population $l$ and the baseline population $k$. Suppose the baseline population is neither $k$ nor $l$, the other strategy is to use the already defined population indicators and test
$$ H_0:\eta_k-\eta_l=0.$$
The Wald statistics for both null hypotheses are identical; see Supplementary Materials for derivations. 

As before, the exact form of the Wald test statistic depends on the genomic region. For an autosomal or X chromosomal PAR SNP,
\begin{equation}\label{sdMAFdiff_A}
    W_{\text{sdMAF diff,A}}^{\scriptscriptstyle{(k,l)}}=\frac{[(\hat p_{f,k} - \hat p_{m,k}) - (\hat p_{f,l} - \hat p_{m,l})]^2}{\splitfrac{\frac{1}{2n_{f,k}}(\hat p_{f,k}(1-\hat p_{f,k})+\hat\delta_{f,k})+\frac{1}{2n_{m,k}}(\hat p_{m,k}(1-\hat p_{m,k})+\hat\delta_{m,k})}{+\frac{1}{2n_{f,l}}(\hat p_{f,l}(1-\hat p_{f,l})+\hat\delta_{f,l})+\frac{1}{2n_{m,l}}(\hat p_{m,l}(1-\hat p_{m,l})+\hat\delta_{m,l})}}\stackrel{H_0}{\sim}\chi_1^2.
\end{equation}
In contrast, for a X chromosomal NPR SNP,
\begin{equation}\label{sdMAFdiff_X}
    W_{\text{sdMAF diff,X}}^{\scriptscriptstyle{(k,l)}}=\frac{[(\hat p_{f,k} - \hat p_{m,k}) - (\hat p_{f,l} - \hat p_{m,l})]^2}{\splitfrac{\frac{1}{2n_{f,k}}(\hat p_{f,k}(1-\hat p_{f,k})+\hat\delta_{f,k})+\frac{1}{n_{m,k}}\hat p_{m,k}(1-\hat p_{m,k})}{+\frac{1}{2n_{f,l}}(\hat p_{f,l}(1-\hat p_{f,l})+\hat\delta_{f,l})+\frac{1}{n_{m,l}}\hat p_{m,l}(1-\hat p_{m,l})}}\stackrel{H_0}{\sim}\chi_1^2.
\end{equation}
Each of the above two test statistics is intuitive as the numerator contains the difference in estimated sdMAF between the two populations, while the denominator contains the estimated variance of the sdMAF difference, where the variance takes different forms for males depending on the genomic region.

\begin{rmk}
The between-population sdMAF tests in \eqref{sdMAFdiff_A} and \eqref{sdMAFdiff_X} can detect the scenario where two populations have similar sdMAF magnitude, but in {\it opposite} directions. 
\end{rmk}

For more than two-population sdMAF comparison, say all $K$ populations without loss of generality, the multi-population between-population sdMAF test can be conducted by testing
$$H_0:\eta_{1}=\dots=\eta_{K-1}=0.$$
For an automsomal or X chromosomal PAR SNP, the Wald statistic is
\begin{equation}\label{sdMAFdiff_mult_A}
\begin{aligned}
    W_{\text{sdMAF diff,A}}^{\scriptscriptstyle{(0,1,\dots,K-1)}} = &  \sum_{k=1}^{K-1} \{[(\hat p_{f,k} - \hat p_{m,k}) - (\hat p_{f,0} - \hat p_{m,0})]^2\frac{\hat U_{k,A}\sum_{l\neq k}\hat U_{l,A}}{\hat U_{k,A}+\sum_{l\neq k}\hat U_{l,A}}\} \\
    - 2 &\sum_{k=1}^{K-1}\sum_{j>k} \{[(\hat p_{f,j} - \hat p_{m,j}) - (\hat p_{f,0} - \hat p_{m,0})][(\hat p_{f,k} - \hat p_{m,k}) - (\hat p_{f,0} - \hat p_{m,0})]\frac{\hat U_{j,A}\hat U_{k,A}}{\sum_{l=0}^{K-1}\hat U_{l,A}}\} \\
    = & \sum_{k=0}^{K-1}(\hat p_{f,k} - \hat p_{m,k})^2\hat U_{k,A}-\frac{(\sum_{k=0}^{K-1}(\hat p_{f,k} - \hat p_{m,k})\hat U_{k,A})^2}{\sum_{k=0}^{K-1}\hat U_{k,A}} \\
    \stackrel{H_0}{\sim} & \chi_{K-1}^2,
\end{aligned}
\end{equation}
where 
$$\hat U_{k,A}=\frac{1}{\frac{1}{2n_{f,k}}(\hat p_{f,k}(1-\hat p_{f,k})+\hat\delta_{f,k})+\frac{1}{2n_{m,k}}(\hat p_{m,k}(1-\hat p_{m,k})+\hat\delta_{m,k})}.$$
By replacing all the subscripts $_A$'s in \eqref{sdMAFdiff_mult_A} with $_X$'s, and define
$$\hat U_{k,X}=\frac{1}{\frac{1}{2n_{f,k}}(\hat p_{f,k}(1-\hat p_{f,k})+\hat\delta_{f,k})+\frac{1}{n_{m,k}}\hat p_{m,k}(1-\hat p_{m,k})},$$  
we obtain $W_{\text{sdMAF diff,X}}^{\scriptscriptstyle{(0,1,\dots,K-1)}}$ for between-population sdMAF comparison, across $K$ populations, for a X chromosomal NPR SNP. 

\begin{rmk}
We were able to simplify the initial complex expression of $W_{\text{sdMAF diff,A}}^{\scriptscriptstyle{(0,1,\dots,K-1)}}$ in \eqref{sdMAFdiff_mult_A} to $\sum_{k=0}^{K-1}(\hat p_{f,k} - \hat p_{m,k})^2\hat U_{k,A}-\frac{(\sum_{k=0}^{K-1}(\hat p_{f,k} - \hat p_{m,k})\hat U_{k,A})^2}{\sum_{k=0}^{K-1}\hat U_{k,A}}$, which is symmetric across the $K$ populations. This is desirable as the test is invariant to the choice of the reference population. The result is also intuitive: If population $k=1$ was the baseline population instead of the initial $k=0$, then using the initial population indicators, we could test $H_0: -\eta_1=0, \eta_2-\eta_1=0, \ldots, \eta_{K-1}-\eta_1=0$. 
\end{rmk}

\section{Empirical Evaluation} \label{sec:3}

We conducted simulation studies to evaluate the proposed multi-population sdMAF tests and between-population sdMAF comparison tests. We focused on sizes of the proposed tests, as a) the existing multi-population sdMAF test is known to be conservative with deflated type I error rate (S1 note of \cite{wang_major_2022}), and b) there is no existing between-population sdMAF comparison test for power comparison. Additionally, the power of each proposed test depends on population-specific sample size, sdMAF effect size and direction in a predictable way, given the closed-form and easy-to-interpret test statistics in \eqref{sdMAF_multtest1_A}, \eqref{sdMAF_multtest1_X}, \eqref{sdMAFdiff_A}, and \eqref{sdMAFdiff_X}. Thus,  we bypassed power evaluation via simulation studies. Instead, we used the application to the 1000 Genomes Project data in Section \ref{sec:4} to demonstrate the practical value of the proposed methods.


\subsection{General simulation set-up} \label{sec:3.1}

To make the simulation realistic, we utilized the high-coverage data of the 1000 Genomes Project \citep{byrska-bishop_high-coverage_2022} that was previously analyzed by \cite{wang_major_2022}. There are 2,504 individuals from the five super-populations. The sample sizes are AFR=661 (319 males:342 females), AMR=347 (170:177), EAS=504 (244:260), EUR=503 (240:263), and SAS=489 (260:229). We included all 2,504 samples in the analysis as in \cite{wang_major_2022}, and regarded the five super-populations as the populations defined in model \eqref{mult_model}.  

We focused on common variants, defined as SNPs with MAF $\geq0.05$ in each of the five super-populations; joint analysis of multiple rare variants is beyond the scope of this work.  In total, 7,915 X chromosomal PAR SNPs and 109,247 X chromosomal NPR SNPs were analyzed. Additionally, 238,367 common autosomal SNPs, all from chromosome 7, were also analyzed; chromosome 7 was chosen as it is the autosome most similar in length to the X chromosome. 

To obtain realistic (autosomal and X chromosomal PAR and NPR) genotype data under the null, we used the observed {\it genotype} frequencies to simulate genotype counts. We note that we simulated genotype counts, instead of allele counts, so that the simulated data maintain any Hardy-Weinberg disequilibrium present in the real data. Additionally, the simulations were super-population-stratified to allow for super-population-specific MAF and HWD. We next describe in detail the simulation methods and results for the two null scenarios: no sdMAF in any of the five super-populations as studied in Section \ref{sec:3.2}, and no difference in sdMAF between populations as studied in Section \ref{sec:3.3}. 

\subsection{Multi-population sdMAF simulation method and results} \label{sec:3.2}

Here we simulated data to evaluate the accuracy of the proposed multi-population sdMAF tests, using \eqref{sdMAF_multtest1_A} for an autosomal or X chromosomal PAR SNP, and \eqref{sdMAF_multtest1_X} for a X chromosomal NPR SNP.  

First, we obtained population-specific genotype frequencies from females, for each of the 355,529 SNPs analyzed (238,367 chromosome 7, and 7,915 PAR and 109,247 NPR SNPs). We let these be the event probabilities for the population-specific multinomial distributions, and we used these to simulate  genotype counts for females for each of the five super-populations in the 1000 Genomes Project. 

Second, for males and each autosomal or PAR SNP, we simulated genotype counts using the population-specific multinomial distributions derived from females. Clearly, this simulation method generates null data with no sdMAF within each of the five super-population groups (i.e.\ the multi-population sdMAF null), while allowing for population-specific MAF and HWD. 

Finally, for males and each PAR SNP, we simulated population-specific hemizygous genotype (i.e.\ allele) counts using binomial distributions, where the corresponding event probabilities are the MAFs in females.   

\begin{figure}[htbp]
\centering
\includegraphics[width=12cm]{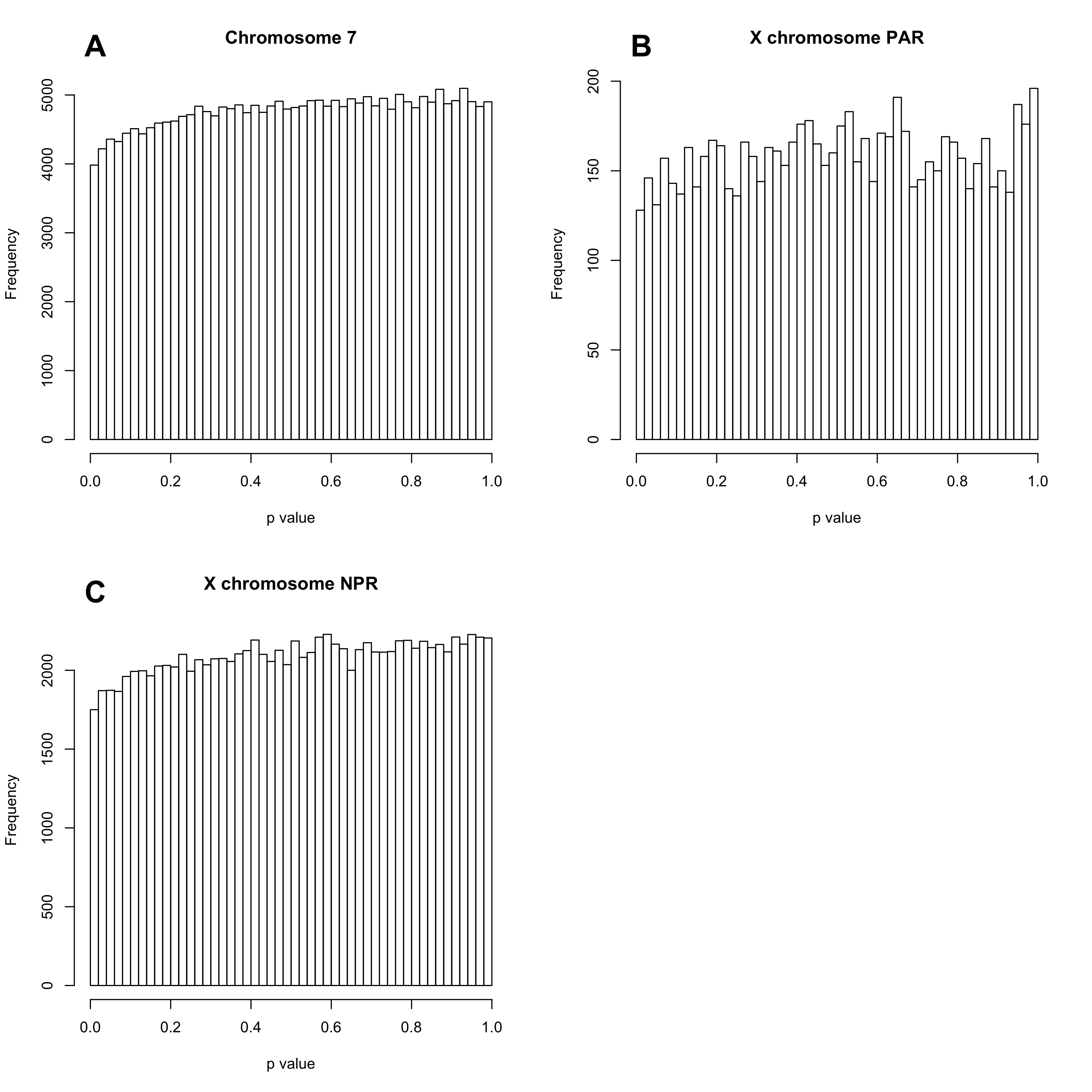} 
\caption[Empirical evaluation of the existing multi-population sdMAF tests]
{{\bf Type I error evaluation of the existing multi-population sdMAF tests (i.e.\ ignoring the population structure), using \eqref{sg_auto} for an autosomal or X chromosomal PAR SNP, and \eqref{sg_npr} for a X chromosomal NPR SNP, using randomized genotype data from the 1000 Genomes Project high-coverage dataset.} The null data were obtained by generating population-stratified genotype counts, respectively for females and males, both based on the observed genotype frequencies in females of the 1000 Genomes Project data; see Section \ref{sec:3.2} for simulation details. The histograms of the p-values are given in (A), (B) and (C), respectively for chromosome 7 (238,367 SNPs), X chromosome PAR (7,915 SNPs) and  and NPR (109,247 SNPs). These empirical results are consistent with the earlier reports that not accounting for population structure leads to  {\it deflated} type I error in sdMAF analysis, in contrast to association analysis.}
\label{EmpEval_1df}
\end{figure}

In addition to the proposed multi-population sdMAF tests, \eqref{sdMAF_multtest1_A} and \eqref{sdMAF_multtest1_X}, for completeness, we also applied the tests \eqref{sg_auto} and \eqref{sg_npr} which ignore the population structure. This is merely to provide empirical evidence to show that unadjusted population structure could lead to {\it deflated} type I error; the work of \cite{wang_major_2022} has provided theoretical justification for the overestimation of variance (through HWD $\delta$) when MAFs differ between populations. Indeed, the three histograms in Figure~\ref{EmpEval_1df} clearly show that, unlike association studies, ignoring population structure results in conservative sdMAF tests for both the autosomal SNPs (Figure~\ref{EmpEval_1df}A), X chromosomal PAR SNPs (Figure~\ref{EmpEval_1df}B) and NPR SNPs (Figure~\ref{EmpEval_1df}C). 

\begin{figure}[htbp]
\centering
\includegraphics[width=12cm]{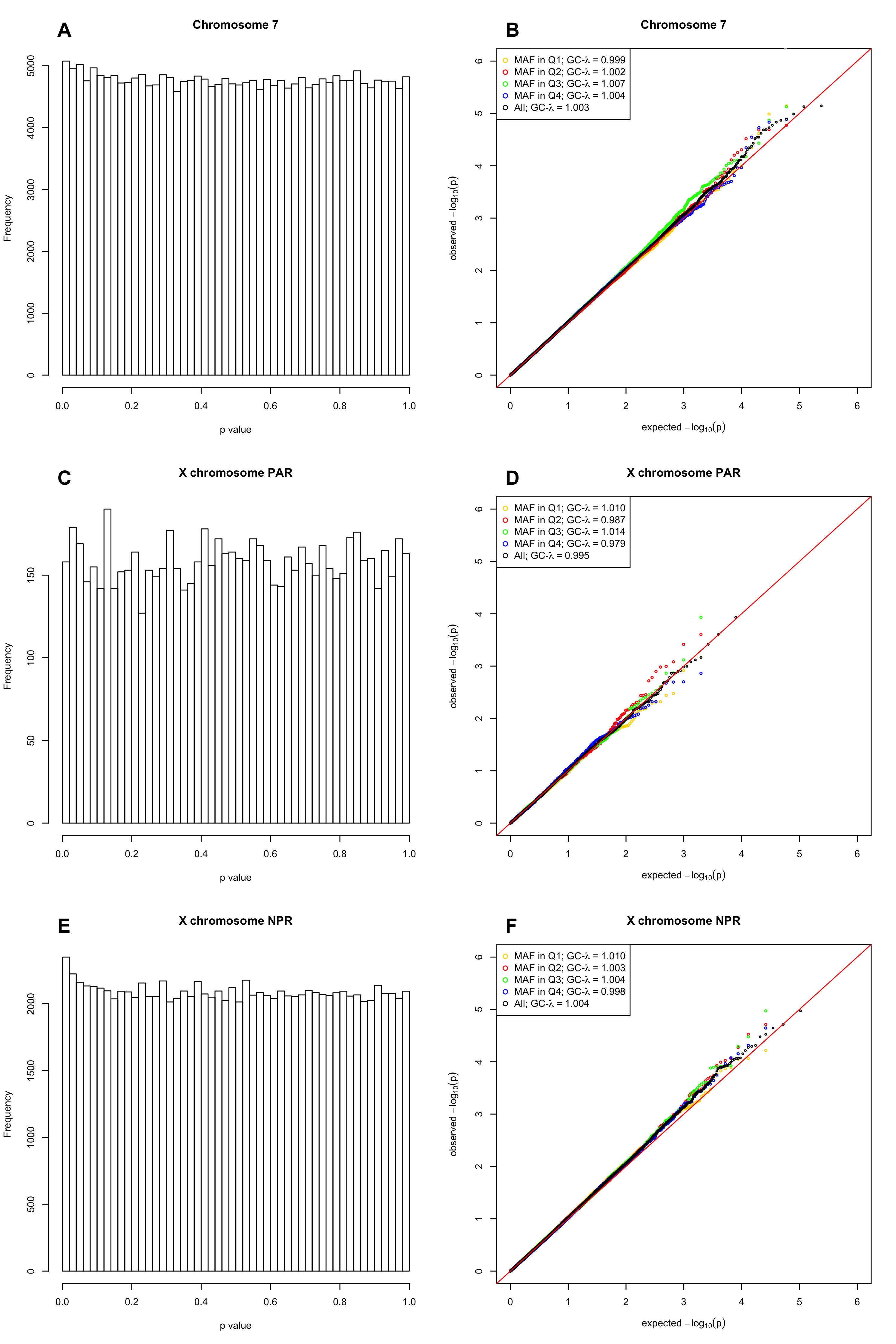} 
\caption[Empirical evaluation of the multi-population sdMAF tests]
{{\bf Type I error evaluation of the proposed multi-population sdMAF tests, using \eqref{sdMAF_multtest1_A} for an autosomal or X chromosomal PAR SNP, and  \eqref{sdMAF_multtest1_X} for a X chromosomal NPR SNP, using randomized genotype data from the 1000 Genomes Project high-coverage dataset.} The null data were obtained by generating population-stratified genotype counts, respectively for females and males, both based on the observed genotype frequencies in females of the 1000 Genomes Project data; see Section \ref{sec:3.2} for simulation details. The empirical distributions of the p-values are given in (A), (C) and (E), respectively for chromosome 7 (238,367 SNPs), X chromosome PAR (7,915 SNPs) and NPR (109,247 SNPs). The corresponding QQ-plots comparing the empirical p-values with Unif(0,1) distribution are given in (B), (D) and (F). The QQ-plots include MAF-stratified QQ-plots, by splitting the SNPs into four equal-sized sets according to the quartiles of the combined MAFs from the whole sample. The quartiles differ slightly between the different genomic regions, but approximately: yellow=(0.05,0.2); red=(0,2,0.3); green=(0.3,0.4); blue=(0,4,0.5); and black for all SNPs analyzed for each genomic region. The corresponding genomic control $\lambda$ values are also provided, suggesting good type I error control.}
\label{EmpEval_multipop}
\end{figure}

Figure~\ref{EmpEval_multipop} shows the empirical p-value distributions of the proposed multi-population sdMAF tests, stratified by the three genomic regions, the autosomal chromosome 7, and the X chromosome PAR and NPR. Figure~\ref{EmpEval_multipop} includes both the histograms (Figures ~\ref{EmpEval_multipop}A, ~\ref{EmpEval_multipop}C and ~\ref{EmpEval_multipop}E), and MAF-stratified QQ-plots ((Figures ~\ref{EmpEval_multipop}B, ~\ref{EmpEval_multipop}D and ~\ref{EmpEval_multipop}F) with their corresponding genomic control $\lambda$ values \citep{devlin1999genomic}.  Results clearly show that the proposed tests can effectively adjust for population structure, resulting in more accurate multi-population sdMAF tests than the existing sdMAF tests; the later is conservative as shown in Figure~\ref{EmpEval_1df}.  

\subsection{Between-population sdMAF comparison simulation method and results} \label{sec:3.3}

Here we simulated data to evaluate the accuracy of the proposed novel between-population sdMAF comparison tests, using \eqref{sdMAFdiff_A} for an autosomal or X chromosomal PAR SNP, and \eqref{sdMAFdiff_X} for a X chromosomal NPR SNP. We focused on evaluating (and applying in Section \ref{sec:4}) the pairwise between-population sdMAF comparison test, as it is practically more interpretable than the multi-population between-population sdMAF comparison test using \eqref{sdMAFdiff_mult_A}. To implement the pairwise between-population sdMAF comparison tests here (and later in the application), we used EUR as the baseline population without loss of generality. 

The null for between-population sdMAF comparison ideally should allow the presence of sdMAF in each population, albeit at the same magnitude and in the same direction. To achieve this, we simulated genotype counts separately for females and males to create sdMAF, but in a population-combined manner to create the null of no population difference in sdMAF. 

First, we obtained genotype frequencies from females  {\it in the whole sample}. We used the corresponding multinomial distribution to simulate genotype counts for all females, for each of the 355,529 SNPs analyzed
(238,367 chromosome 7, and 7,915 PAR and 109,247 NPR SNPs as in Section \ref{sec:3.2} above). 

Second, we similarly simulated genotype counts for males for each of the 238,367 chromosome 7 and 7,915 PAR SNPs analyzed. That is, we obtained genotype frequencies of males  in the whole sample. We used the corresponding multinomial distributions to simulate genotype counts for all males for these autosomal and PAR SNPs.

Lastly, for each of 109,247 NPR SNPs analyzed, we obtained hemizygous genotype (i.e.\ allele) frequencies from males in the whole sample. We then used the corresponding binomial distributions to simulate the hemizygous genotype (i.e.\ allele) counts for all males for these NPR SNPs.    

\begin{figure}[htbp]
\centering
\includegraphics[width=12cm]{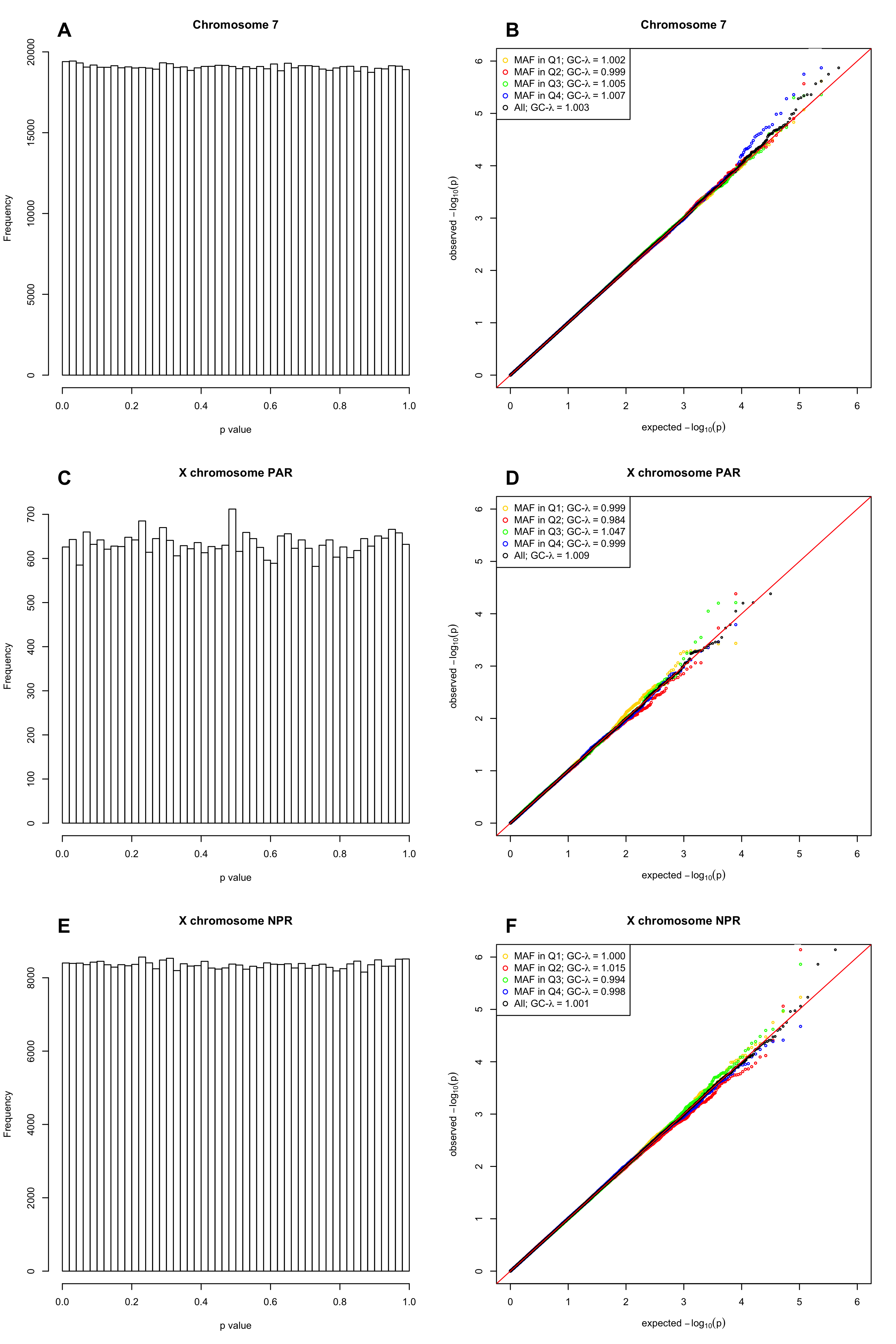} 
\caption[Empirical evaluation of the multi-population sdMAF tests]
{{\bf Type I error evaluation of the proposed between-population sdMAF comparison tests, using \eqref{sdMAFdiff_A} for an autosomal or X chromosomal PAR SNP, and \eqref{sdMAFdiff_X} for a X chromosomal NPR SNP, using randomized genotype data of the 1000 Genomes Project high-coverage dataset.} The null data were obtained by generating population-combined genotype counts, separately, for females and males based on the sex-stratified observed genotype frequencies of the 1000 Genomes Project data; see Section \ref{sec:3.3} for simulation details. The empirical distributions of the p-values are given in (A), (C) and (E), respectively for chromosome 7 (238,367 SNPs), X chromosome PAR (7,915 SNPs) and NPR (109,247 SNPs). The corresponding QQ-plots comparing the empirical p-values with Unif(0,1) distribution are given in (B), (D) and (F). The QQ-plots include MAF-stratified QQ-plots, by splitting the SNPs into four equal-sized sets according to the quartiles of the combined MAFs from the whole sample. The quartiles differ slightly between the different genomic regions, but approximately: yellow=(0.05,0.2); red=(0,2,0.3); green=(0.3,0.4); blue=(0,4,0.5); and black for all SNPs analyzed for each genomic region. The corresponding genomic control $\lambda$ values are also provided, suggesting good type I error control. In this simulation study, the EUR population was chosen to be the baseline population, and all pairwise sdMAF comparison testing results were presented together for simplicity but without loss of generality.}
\label{EmpEval_betweenpop}
\end{figure}

Figure~\ref{EmpEval_betweenpop} shows the empirical p-value distributions of the proposed pairwise between-population sdMAF tests, stratified by the three genomic regions, the autosomal chromosome 7, and the X chromosome PAR and NPR. Similar to Figure~\ref{EmpEval_multipop} in Section \ref{sec:3.2}, Figure~\ref{EmpEval_betweenpop}  also includes both histograms (Figures ~\ref{EmpEval_betweenpop}A, ~\ref{EmpEval_betweenpop}C and ~\ref{EmpEval_betweenpop}E), and MAF-stratified QQ-plots (Figures ~\ref{EmpEval_betweenpop}B, ~\ref{EmpEval_betweenpop}D and ~\ref{EmpEval_betweenpop}F) with their corresponding genomic control $\lambda$ values.  Results show that the new between-population sdMAF comparison tests are accurate in the presence of sdMAF, as long as there are no sdMAF differences between populations. 

\section{Application to the 1000 Genomes Project high-coverage data} \label{sec:4} 

We applied the different tests to the observed high-coverage data of the 1000 Genomes Project \citep{byrska-bishop_high-coverage_2022}. The data were already described in Section \ref{sec:3.1} for simulation studies, where we randomized the observed data. Briefly, we analyzed 2,504 individuals from five super-populations (AFR, AMR, EAS, EUR, and SAS) with similar population- and sex-stratified sample sizes, and 355,529 common SNPs (238,367 chromosome 7, and 7,915 PAR and 109,247 NPR X chromosomal SNPs).

Similar to the simulation studies, we first applied the proposed multi-population sdMAF tests (using \eqref{sdMAF_multtest1_A} for autosomal and PAR SNPs, and \eqref{sdMAF_multtest1_X} for NPR SNPs), and compared the results with the existing tests (\eqref{sg_auto} and \eqref{sg_npr}). We then applied the new pairwise between-population sdMAF comparison tests (\eqref{sdMAFdiff_A} and \eqref{sdMAFdiff_X}). We used the genome-wide significance association threshold of 5e-8 \citep{dudbridge2008estimation} to declare significance for our sdMAF evaluations. Below we summarize the results, separately for the X chromosome and chromosome 7, first for the multi-population sdMAF tests, then for the pairwise between-population sdMAF comparison tests.

\subsection{The multi-population sdMAF testing results for the X chromosome} \label{sec:4.1} 

\begin{figure}[htbp]
\centering
\includegraphics[width=13cm]{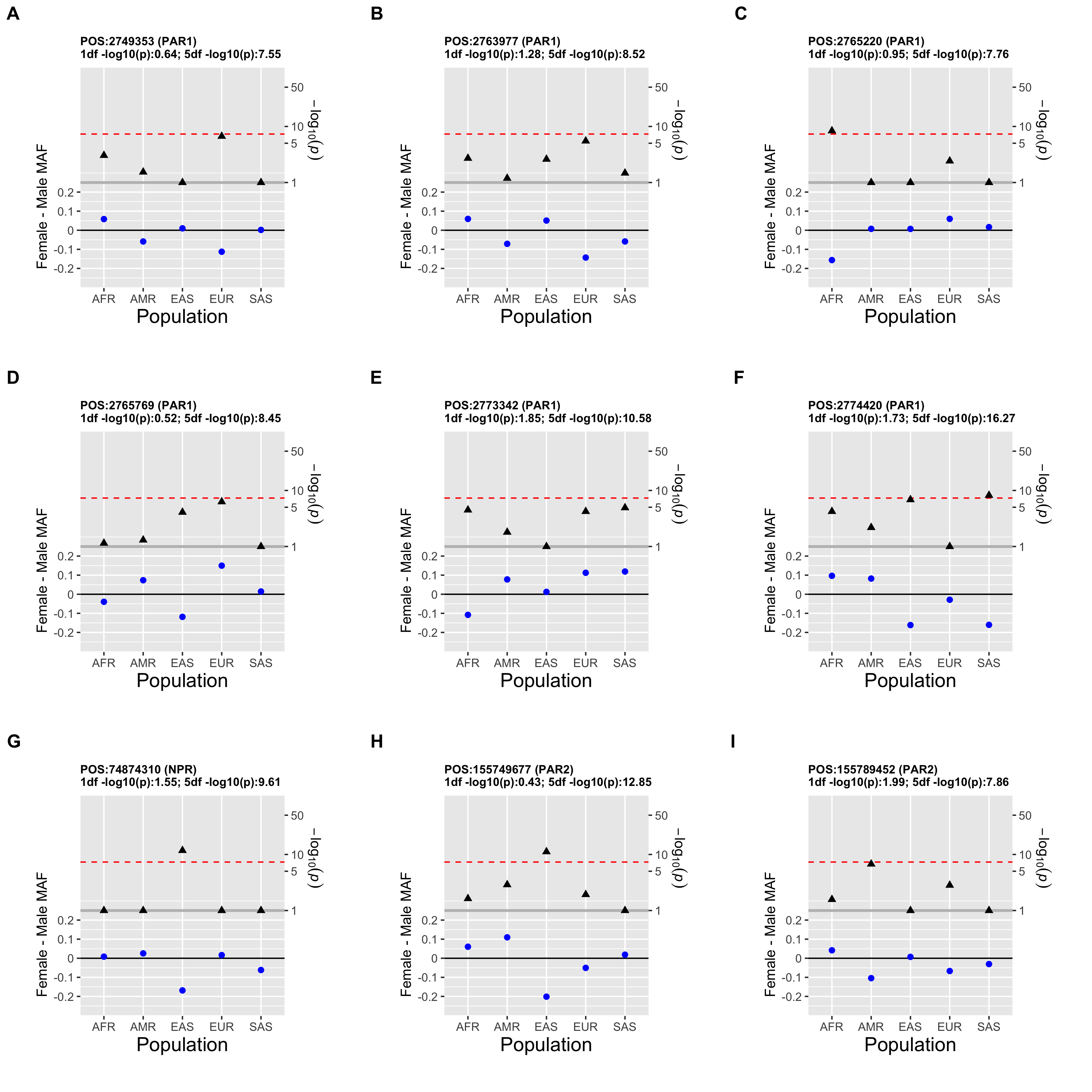}
\caption[SNPs whose p-values were genome-wide significant from the multi-population sdMAF test only]
{{\bf A subset of the 76 X chromosomal SNPs that are genome-wide significant based on the proposed 5 df multi-population sdMAF method but not the existing 1 df whole-sample sdMAF method, in the 1000 Genomes Project high-coverage data application.}  These nine SNPs have the biggest difference in -log10 p-values between the two sdMAF tests, and they are ordered from figure A to figure I based on the genomic position (POS Build GRCh38). In each figure,  the lower panel (blue dots, left y-axis) shows the population-specific sdMAF magnitude and sign for each of the five super-populations; the black horizontal line marks zero. The upper panel (black triangles, right y-axis) shows the corresponding population-specific sdMAF p-value on the -log10 scale;
the red horizontal line marks the genome-wide significance level of -log10(5e-8)=7.3. 
For better visualization, the -log10 p-values are truncated at 1 (i.e.\ sdMAF p-values greater than 0.1 are shown as 0.1).}
\label{fig:1dfnonsig_5dfsignif}
\end{figure}

The proposed 5 df multi-population sdMAF testing method is omnibus, and sdMAF directions may differ between populations, thus new discoveries were expected as compared with the existing 1 df sdMAF method. Indeed, in total there are 76 genome-wide significant X chromosomal (both PAR and NPR) SNPs with sdMAF based on the proposed tests, but were missed by the existing tests. Figure~\ref{fig:1dfnonsig_5dfsignif} shows a subset of these 76 SNPs, ordered by the genomic position. 

Consistent with the expectation based on analytical results in Section \ref{sec:3} (e.g.\ Remark 2), these SNPs have sdMAF directions varying between populations.
For example, for PAR1 SNP in POS:2774420 in Figure~\ref{fig:1dfnonsig_5dfsignif}F, females have higher MAF than males in AFR and AMR, but for EAS, EUR and SAS, females have lower MAF than males. In this scenario, the proposed 5 df sdMAF testing method is robust as it aggregates the population-specific sdMAF estimates quadratically, while the existing 1 df testing method loses power as it does so linearly.  As noted earlier, another scenario that the existing 1 df testing method can lose power is when MAFs differ significantly between populations, which leads to overestimation of HWD $\delta$ in the variance, thus resulting in a conservative test. 

\begin{figure}[htbp]
\centering
\includegraphics[width=13cm]{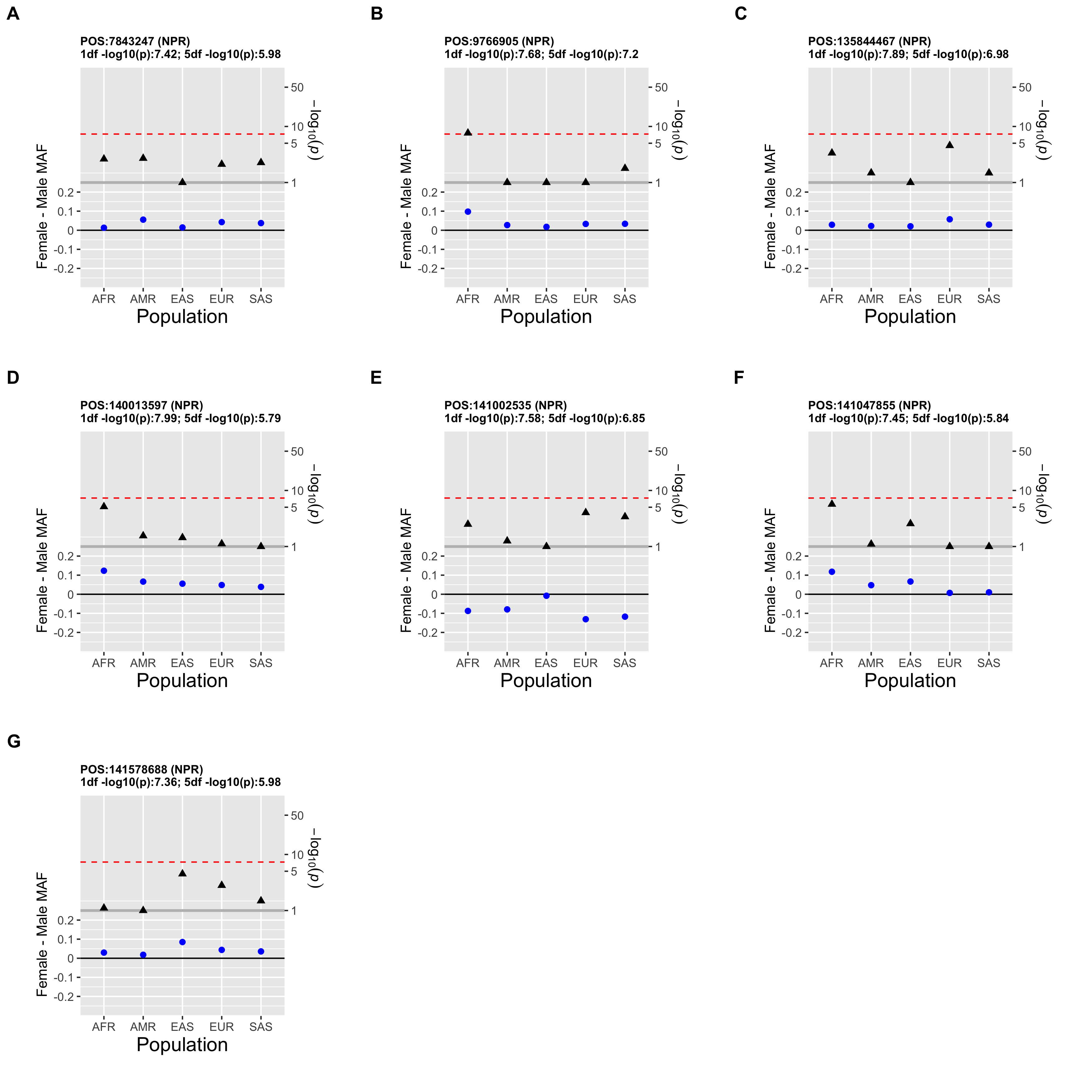}
\caption[SNPs whose p-values were genome-wide significant from the whole-sample 1 df sdMAF test only]
{{\bf All the seven X chromosomal SNPs that are genome-wide significant based on existing 1 df whole-sample sdMAF method but missed by the proposed 5 df multi-population sdMAF method, in the 1000 Genomes Project high-coverage data application.}  The SNPs are ordered from figure A to figure I based on the genomic position (POS Build GRCh38. In each figure,  the lower panel (blue dots, left y-axis) shows the population-specific sdMAF magnitude and sign for each of the five super-populations; the black horizontal line marks zero. The upper panel (black triangles, right y-axis) shows the corresponding population-specific sdMAF p-value on the -log10 scale;
the red horizontal line marks the genome-wide significance level of -log10(5e-8)=7.3. 
For better visualization, the -log10 p-values are truncated at 1 (i.e.\ sdMAF p-values greater than 0.1 are shown as 0.1).}
\label{fig:1dfsignif_5dfnonsig}
\end{figure}

When sdMAF directions (and magnitude) are similar across populations, the existing 1 df multi-population sdMAF method is more powerful than the proposed 5 df method (Figure~\ref{fig:1dfsignif_5dfnonsig}). This is expected, but two features of Figure~\ref{fig:1dfsignif_5dfnonsig} are worth commenting on. First, the total number of SNPs missed by the proposed multi-population sdMAF method is seven, as compared with 76 that were missed by the existing method (nine of which are shown in Figure~\ref{fig:1dfnonsig_5dfsignif}). Second, compared with results in Figure~\ref{fig:1dfnonsig_5dfsignif}, for all the seven SNPs missed by the proposed method in Figure~\ref{fig:1dfsignif_5dfnonsig}, the sdMAF p-values of the two methods are comparable. For example, the smallest p-value of the seven SNPs is -log10(1e-8)=7.99 based on the existing 1 df method (Figure~\ref{fig:1dfsignif_5dfnonsig}D). While the proposed 5 df test was not statistically significant, its sdMAF testing result is -log10(1.6e-6)=5.79. In contrast, for example, Figure~\ref{fig:1dfnonsig_5dfsignif}F shows that a significant SNP detected by the proposed method has -log10(5.3e-17)=16.27, while the existing method ranks this SNP very low with -log10(1.85e-2)=1.73. 

\begin{figure}[htbp]
\centering
\includegraphics[width=13cm]{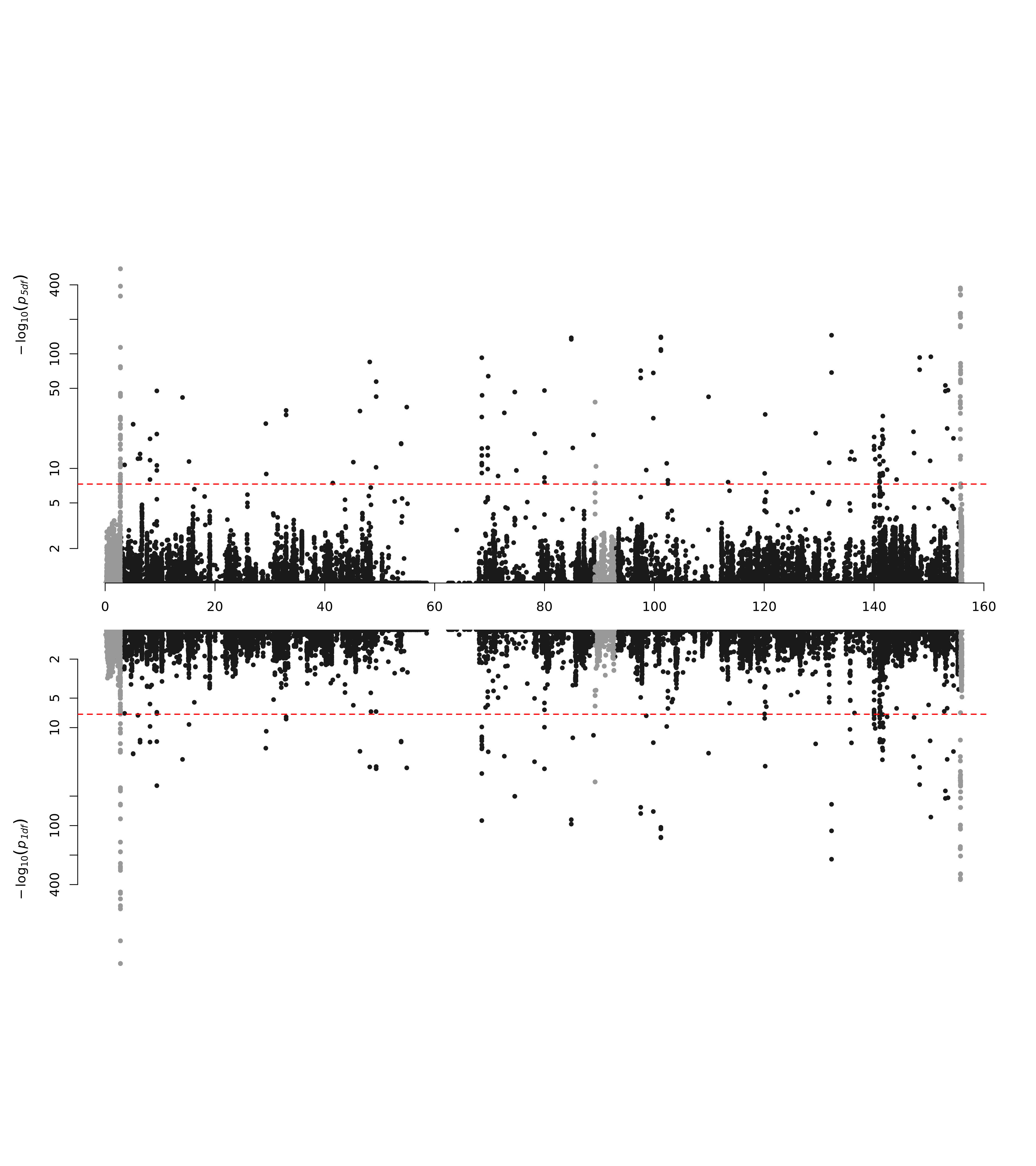}
\caption[Miami plot comparing p-values between the one-degree of freedom sdMAF test applied to the whole sample and the five-degree of freedom multi-population sdMAF test]
{{\bf Miami plot comparing the multi-population sdMAF p-values between the proposed 5 df method (upper panel) and the existing 1 df method (lower panel), in the 1000 Genomes Project high-coverage data application.}  The X-axis is the physical position in GRCh38 in Mb. SNPs in PAR1, PAR2 and PAR3 are plotted in grey, with PAR1 at the beginning of the X chromosome starting at 0 Mb, PAR3 around 90 Mb, and PAR2 at the end.Y-axis is -log10(p-values) and p-values >0.1 are plotted as 0.1 (1 on -log10 scale) for better visualization. The dashed red line represents 5e-8 (7.3 on the -log10 scale).}
\label{fig:1df_vs_5df_Miami}
\end{figure}

\begin{figure}[htbp]
\centering
\includegraphics[width=13cm]{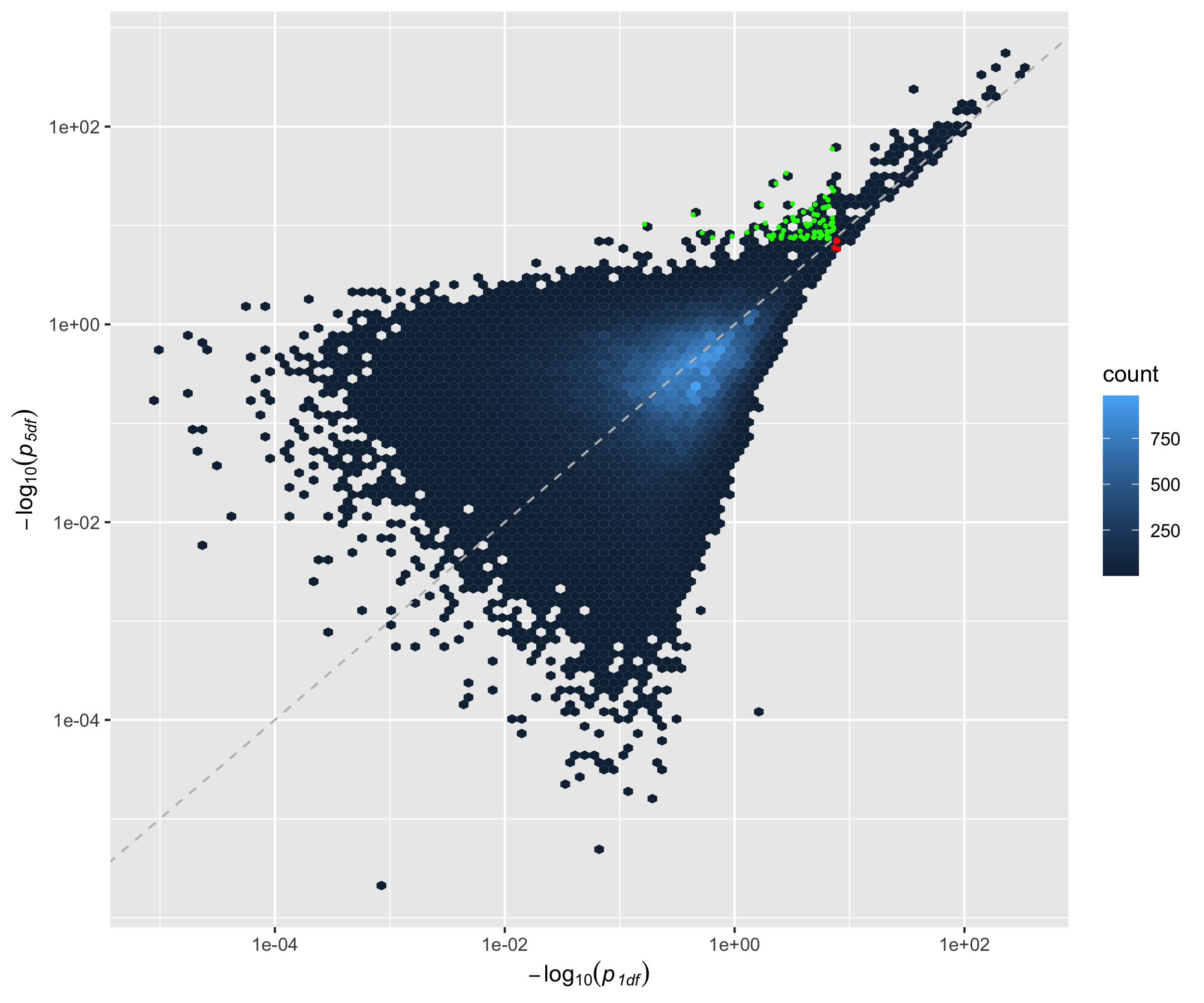}
\caption[P-P hexbin plot comparing p-values between the one-degree of freedom sdMAF test applied to the whole sample and the five-degree of freedom multi-population sdMAF test]
{{\bf P-P hexbin plot comparing the multi-population sdMAF -log10 (p-values) between the proposed 5 df method (Y-axis) and the existing 1 df method (X-axis), in the 1000 Genomes Project high-coverage data application.}  Green dots are genome-wide significant sdMAF SNPs based on the proposed method but non-significant based on the existing method; Red dots are genome-wide significant sdMAF SNPs based on the existing method but non-significant based on the proposed method.}
\label{fig:1df_vs_5df_PPplot}
\end{figure}

Across the whole X chromosome, Figure~\ref{fig:1df_vs_5df_Miami} (the so-called Miami plot) contrasts the proposed method (upper panel) with the existing approach (lower panel). There are grey areas representing the PAR1, PAR3 and PAR2 regions, from left to right. 
To further demonstrate the improved power of the proposed method, Figure~\ref{fig:1df_vs_5df_PPplot} shows the P-P  hexbin plot. It is clear that a) the proposed mutli-population method can identify SNPs with significant sdMAF that are completely missed by the existing method, with several orders of magnitude difference in p-values, and b) the proposed method may miss some signals but still offers comparable sdMAF evidence.   

\subsection{The pairwise between-population sdMAF comparison testing results of the X chromosome} \label{sec:4.2} 

\begin{figure}[htbp]
\centering
\includegraphics[width=10cm]{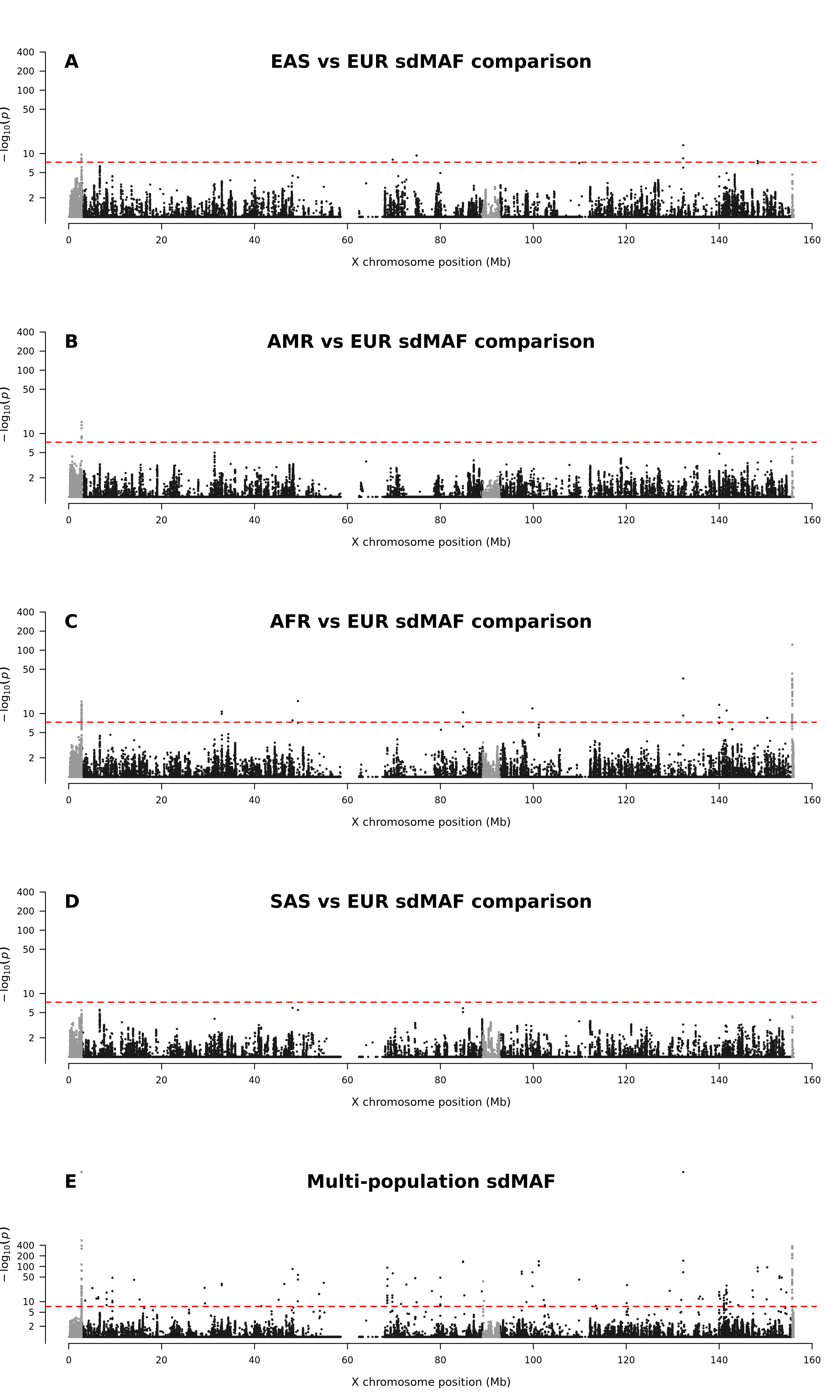}
\caption[sdMAF comparison test and multi-population sdMAF test results of X Chromosome from 1000 Genomes Project high-coverage dataset]
{{\bf The proposed pairwise between-population sdMAF comparison testing X chromosomal results in the 1000 Genomes Project high-coverage data application.}  To assist results interpretation, figure E shows the proposed multi-population sdMAF testing results already shown in the upper panel of Figure~\ref{fig:1df_vs_5df_Miami}. SNPs in PAR1, PAR2 and PAR3 are plotted in grey, with PAR3 located around 90 Mb. Y-axis is -log10(p-values) and p-values >0.1 are plotted as 0.1 (1 on -log10 scale) for better visualization. The dashed red line represents 5e-8 (7.3 on the -log10 scale).}
\label{fig:sdMAFcomparison_chrX}
\end{figure}

Figures~\ref{fig:sdMAFcomparison_chrX}A-D provide the pairwise between-population sdMAF {\it comparison} test, where the EUR was the chosen baseline population. Recall that this tests for differences in sdMAF between populations, while allowing for sdMAF to be present in each of the five populations. To assist results interpretation, Figure~\ref{fig:sdMAFcomparison_chrX}E provides the multi-population sdMAF testing results from Section \ref{sec:4.1} above. 

Two notable features stand out in Figure~\ref{fig:sdMAFcomparison_chrX}. First, most of the NPR SNPs with genome-wide significant multi-population sdMAF (Figure~\ref{fig:sdMAFcomparison_chrX}E) do not have significant between-population sdMAF (Figures~\ref{fig:sdMAFcomparison_chrX}A-D). Second, in contrast, many of the PAR1 and PAR2 SNPs with genome-wide significant multi-population sdMAF have evidence for between-population difference in sdMAF, particularly between the AFR and EUR populations. 


\subsection{The chromosome 7 results} \label{sec:4.3} 

\begin{figure}[htbp]
\centering
\includegraphics[width=10cm]{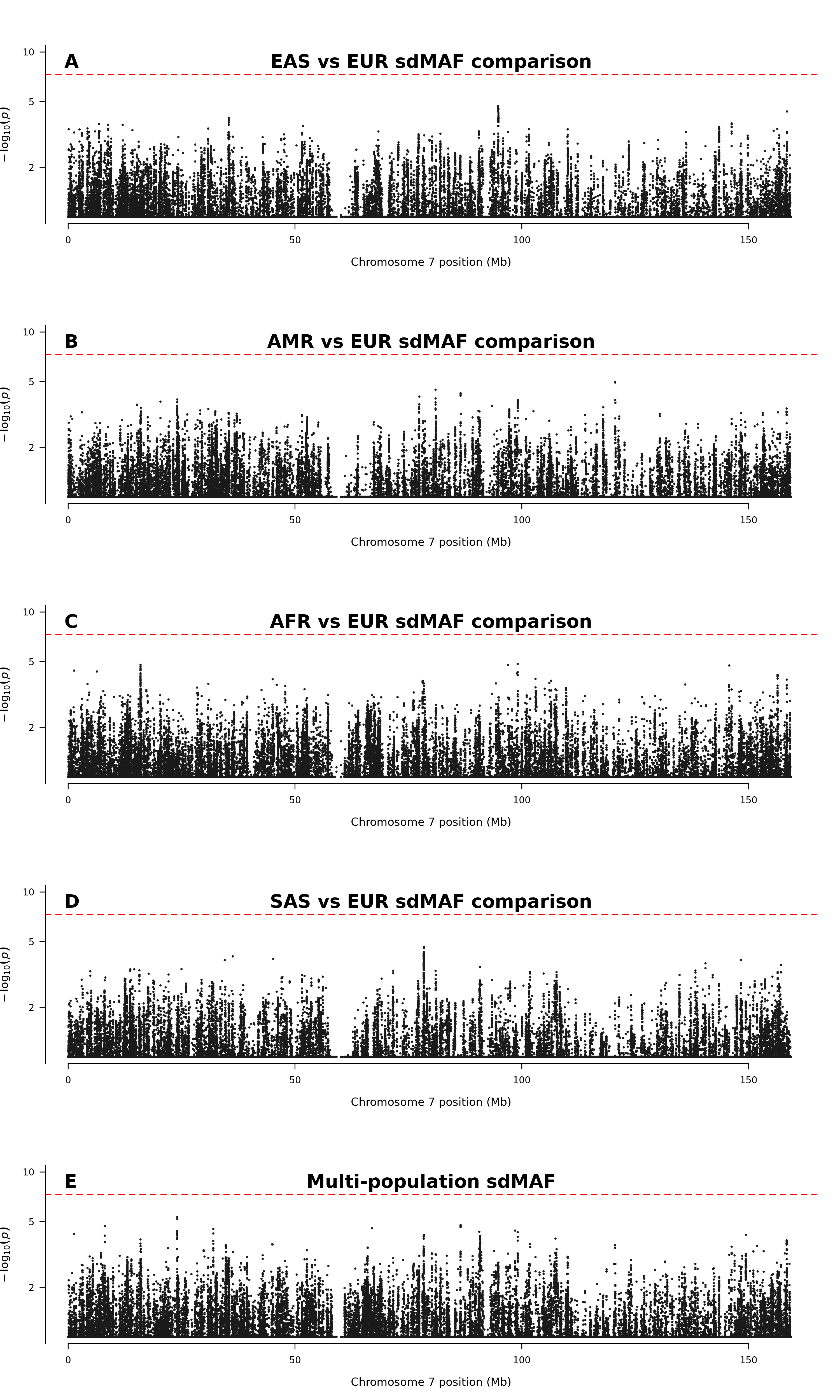}
\caption[sdMAF comparison test and multi-population sdMAF test results of Chromosome 7 from 1000 Genomes Project high-coverage dataset]
{{\bf The proposed pairwise between-population sdMAF comparison testing chromosome 7 results in the 1000 Genomes Project high-coverage data application.}  To assist results interpretation, figure E shows the proposed multi-population sdMAF testing results.  Y-axis is -log10(p-values) and p-values >0.1 are plotted as 0.1 (1 on -log10 scale) for better visualization. The dashed red line represents 5e-8 (7.3 on the -log10 scale).}
\label{fig:sdMAFcomparison_chr7}
\end{figure}

Figure~\ref{fig:sdMAFcomparison_chr7} shows the 
sdMAF testing results for chromosome 7. Similar to Figure~\ref{fig:sdMAFcomparison_chrX} for the X chromosome, Figures~\ref{fig:sdMAFcomparison_chr7}A-D provide the pairwise between-population sdMAF {\it comparison} test, where the EUR was the chosen baseline population, and Figure~\ref{fig:sdMAFcomparison_chr7}E shows the multi-population sdMAF testing results. 

In contrast to the X chromosomal results, there were no chromosome 7 SNPs with significant sdMAF. This is consistent the results in \cite{wang_major_2022}, where only one significant SNP was found in the earlier version (phase 3 data) of the 1000 Genomes Project. Additionally, there were no chromosome 7 SNPs with significant pairwise between-population sdMAF differences. 

\section{Discussion} \label{sec:5}

Compared with the existing conservative methods to detect sdMAF across multiple populations, the proposed methods identified 76 novel X chromosomal SNPs with genome-wide significant sdMAF in the high-coverage data of the 1000 Genomes Project \citep{byrska-bishop_high-coverage_2022}, jointly analyzing samples from all five-super populations. The gain of power is due to a) correctly modeling Hardy-Weinberg disequilibrium in samples from diverse ancestral groups, and b) recognizing that sdMAF directions may differ between populations.  Additionally, the proposed retrospective regression-based testing framework provides a novel between-population sdMAF comparison test that can formally evaluate population differences in sdMAF. The proposed test identified ancestral differences in sdMAF, particularly at the non-pseudoautosomal regions of the X chromosome. 

In comparison, we report no significant results on  chromosome 7, the autosome with genomic length most similar to that of the X chromosome. This is consistent with the report by \cite{wang_major_2022}, who analyzed the phase 3 (instead of the high-coverage) data of the 1000 Genomes Project and identified only one chromosome 7 SNP with significant sdMAF.  However, our results call for future discussions about the recent reports that there is ``a widespread sex-differential participation bias" \citep{pirastu2021genetic}, identified through an {\it autosome-only} association study of sex as a binary outcome (i.e.\ GWAS of sex).  First, we note that statistically speaking, GWAS of sex is examining sdMAF.  Second, \cite{pirastu2021genetic} used data from a commercial direct-to consumer genetic testing company (23andMe), which likely has strong participation bias. Third, \cite{pirastu2021genetic} noted that 55\% of their significant findings are likely results of genotyping errors. Finally, we note that the X chromosome was omitted from their study. 

Among the proposed two testing frameworks, both the multi-population sdMAF tests (Figure~\ref{EmpEval_multipop}) and the pairwise between-population comparison tests (Figure~\ref{EmpEval_betweenpop}) have good type I error control. However, the between-population comparison tests in  Figure~\ref{EmpEval_betweenpop} appear to behave even better than the multi-population sdMAF tests in Figure~\ref{EmpEval_multipop}. This is expected, because the pairwise between-population sdMAF comparison tests are 1 df, while the multi-population sdMAF tests are 5 df. To this end, we caution against applying the proposed multi-population sdMAF tests to data of small sample sizes and/or SNPs with low minor allele frequencies. Even for the 1 df between-population sdMAF comparison tests, we applied them only to common variants (MAF >5\% in each population).  Thus, it is a future research interest to study rare variants from the sdMAF perspective.

\begin{figure}[htbp]
\centering
\includegraphics[width=10cm]{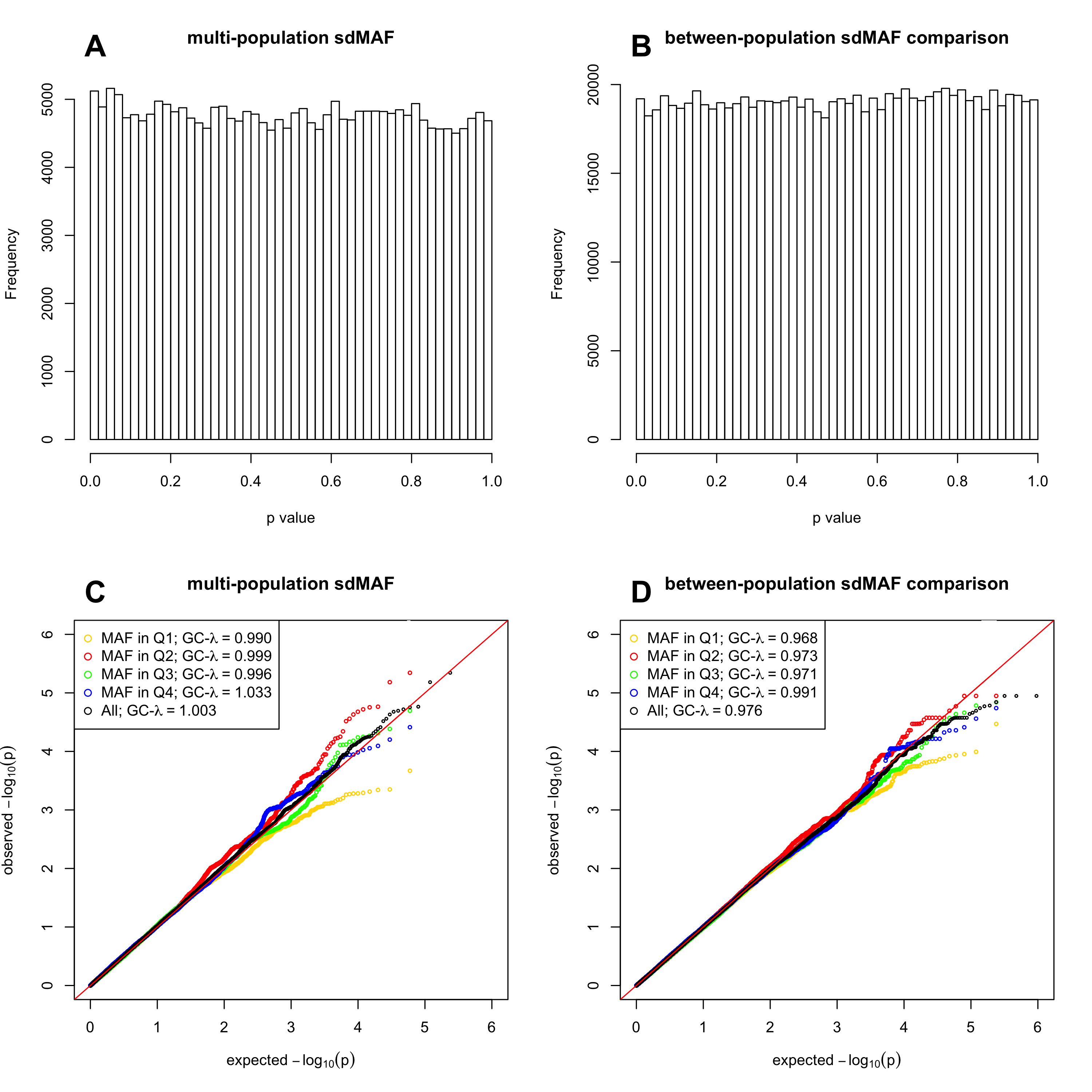}
\caption[sdMAF comparison test and multi-population sdMAF test results of Chromosome 7 from 1000 Genomes Project high-coverage dataset]
{{\bf The histograms and QQ-plots of the chromosome 7 results in the 1000 Genomes Project high-coverage data application.} As we do not expect the autosomes to have significant amount of sdMAF, we expect the p-values to be approximately Unif(0,1) distributed for both the multi-population sdMAF test (A and C) and the between-population sdMAF comparison test (B and D).}
\label{EmpEval_chr7}
\end{figure}

In our 1000 Genomes Project application, we used the five super-populations as the ancestry groups. It is known that additional population structure exists in each of the super-populations \citep{byrska-bishop_high-coverage_2022}. Unlike in phenotype-genotype association studies \citep{devlin1999genomic}, unadjusted population structure {\it decreases} type I error of sdMAF tests, as shown in \cite{wang_major_2022} and confirmed in Figure~\ref{EmpEval_1df}. Our simulation studies suggest that adjusting for the super-population structure is sufficient to achieve accurate type I error control (Figures~\ref{EmpEval_multipop} and \ref{EmpEval_betweenpop}).  As we do not expect the autosomes to have significant amount of sdMAF, results from chromosome 7 thus provide additional evidence for the accuracy of the proposed methods (Figure~\ref{EmpEval_chr7}). Nevertheless, it is of future research interest to examine the effects on sdMAF testing when including principal components \citep{price2006principal} in the proposed model \eqref{mult_model}. Additionally, although our 1000 Genomes Project application did not include other covariates such as age, they can be readily incorporated into the regression model if available. 

Finally, although we only analyzed a set of independent individuals, extension to analyzing related individuals, in principle, can be developed. This is because the proposed tests were derived from regression, for which existing techniques for dependent samples can be leveraged to study sdMAF. 



\section{Resources} \label{sec:6}

High coverage phased data of the 1000 Genomes Project: \url{http://ftp.1000genomes.ebi.ac.uk/vol1/ftp/data_collections/1000G_2504_high_coverage/working/20201028_3202_phased}

Related code: \url{https://github.com/ZhongWang99/sdMAF-via-regression}

\section{Acknowledgement} 
This research was funded by the Canadian Institutes of Health Research (CIHR, PJT-180460), Natural Sciences and Engineering Research Council of Canada (NSERC, RGPIN-04934), and a University of Toronto Data Sciences Institute (DSI) Catalyst Grant. 

\clearpage
\bibliography{AoAS_cite}

\end{document}


\maketitle

\section{Notations}
Denote $G_{i,m}$ as the genotype coding of the $i$th sample from the male group and $G_{i,f}$ as the genotype coding of the $i$th sample from the female group. Denote $n_m$ and $n_f$ as the sample sizes for males and females respectively. Additionally, we let $G_{i,m,k}$, $G_{i,f,k}$, $n_{m,k}$ and $n_{f,k}$ be the corresponding quantities specifically for super-population $k$.

\section{Derivations of (2.6) and (2.7)}
The full log-likelihood $\iota(G)$ can be written as
\begin{equation}
    \iota(G) = -(n_m+n_f)\log(\sqrt{2\pi})-n_m\log\sigma_m - n_f\log\sigma_f - \frac{\sum_{i=1}^{n_m}(G_{i,m}-\alpha)^2}{2\sigma_m^2} - \frac{\sum_{i=1}^{n_f}(G_{i,f}-\alpha-\gamma)^2}{2\sigma_f^2},
\end{equation}
and the first derivatives are
\begin{equation}
    \begin{aligned}
        \frac{\partial\iota}{\partial\sigma_m} & = -\frac{n_m}{\sigma_m} + \frac{\sum_{i=1}^{n_m}(G_{i,m}-\alpha)^2}{\sigma_m^3}, \\
        \frac{\partial\iota}{\partial\sigma_f} & = -\frac{n_f}{\sigma_f} + \frac{\sum_{i=1}^{n_f}(G_{i,f}-\alpha-\gamma)^2}{\sigma_f^3}, \\
        \frac{\partial\iota}{\partial\alpha} & = \frac{\sum_{i=1}^{n_m}(G_{i,m}-\alpha)}{\sigma_m^2} + \frac{\sum_{i=1}^{n_f}(G_{i,f}-\alpha-\gamma)}{\sigma_f^2}, \\
        \frac{\partial\iota}{\partial\gamma} & = \frac{\sum_{i=1}^{n_f}(G_{i,f}-\alpha-\gamma)}{\sigma_f^2},
    \end{aligned}
\end{equation}
by which we can easily obtain the MLEs of the parameters.

Obviously,
\begin{equation}
E\left[\frac{\partial^2\iota}{\partial\alpha\partial\sigma_m}\right] = E\left[\frac{\partial^2\iota}{\partial\alpha\partial\sigma_f}\right] =
E\left[\frac{\partial^2\iota}{\partial\gamma\partial\sigma_m}\right] =
E\left[\frac{\partial^2\iota}{\partial\gamma\partial\sigma_f}\right] = 0;
\end{equation}
meanwhile,
\begin{equation}
E\left[-\frac{\partial^2\iota}{\partial\alpha^2}\right] = \frac{n_m}{\sigma_m^2} + \frac{n_f}{\sigma_f^2}, E\left[-\frac{\partial^2\iota}{\partial\gamma^2}\right] = \frac{n_f}{\sigma_f^2}, E\left[-\frac{\partial^2\iota}{\partial\gamma\partial\alpha}\right] = \frac{n_f}{\sigma_f^2}.
\end{equation}

When testing $H_0: R(\theta)=0$, the Wald statistic is given by $W_n = R(\hat\theta)^T\{C(\hat\theta)^TI_n^{-1}(\hat\theta)C(\hat\theta)\}^{-1}R(\hat\theta)$, where $C(\theta)=\partial R(\theta)/\partial\theta$. In our case, $R(\theta)=\gamma$, and since the full Fisher information matrix is a block diagonal matrix, we only require the part of matrix related to $(\alpha,\gamma)$. The inverse of this two by two Fisher matrix reads as
\begin{equation}
I_n^{-1}(\hat\alpha,\hat\gamma) = \begin{pmatrix}
\frac{\hat\sigma_m^2}{n_m} & -\frac{\hat\sigma_m^2}{n_m} \\
-\frac{\hat\sigma_m^2}{n_m} & \frac{\hat\sigma_f^2}{n_f} + \frac{\hat\sigma_m^2}{n_m}.
\end{pmatrix}
\end{equation}
Therefore, the expression of the Wald statistic is
\begin{equation}
    W_n = \frac{\hat\gamma^2}{\frac{\hat\sigma_f^2}{n_f} + \frac{\hat\sigma_m^2}{n_m}} = \frac{4(\hat p_f - \hat p_m)^2}{\frac{\hat\sigma_f^2}{n_f} + \frac{\hat\sigma_m^2}{n_m}}.
\end{equation}
The expressions of $\hat p_f$, $\hat p_m$, $\hat\sigma_f^2$ and $\hat\sigma_m^2$ are already given in the manuscript.

\section{Derivations of (2.9)}
The model is given in (2.8) from the manuscript. The full log-likelihood $\iota(G)$ can be written as
\begin{equation}
\begin{aligned}
    \iota(G) = & - \sum_{k=0}^{K-1}(n_{m,k}+n_{f,k})\log(\sqrt{2\pi})-\sum_{k=0}^{K-1}n_{m,k}\log\sigma_{m,k} - \sum_{k=0}^{K-1}n_{f,k}\log\sigma_{f,k} \\ 
    & - \sum_{k=1}^{K-1}\frac{\sum_{i=1}^{n_{m,k}}(G_{i,m,k}-\alpha-\gamma_k)^2}{2\sigma_{m,k}^2} - \sum_{k=1}^{K-1}\frac{\sum_{i=1}^{n_{f,k}}(G_{i,f,k}-\alpha-\gamma-\gamma_k-\eta_k)^2}{2\sigma_{f,k}^2} \\
    & - \frac{\sum_{i=1}^{n_{m,0}}(G_{i,m,0}-\alpha)^2}{2\sigma_{m,0}^2} - \frac{\sum_{i=1}^{n_{f,0}}(G_{i,f,0}-\alpha-\gamma)^2}{2\sigma_{f,0}^2},
\end{aligned}
\end{equation}
and the first derivatives are
\begin{equation}
    \begin{aligned}
        \frac{\partial\iota}{\partial\sigma_{m,k}} & = -\frac{n_{m,k}}{\sigma_{m,k}} + \frac{\sum_{i=1}^{n_{m,k}}(G_{i,m,k}-\alpha-\gamma_k)^2}{\sigma_{m,k}^3}, \\
        \frac{\partial\iota}{\partial\sigma_{f,k}} & = -\frac{n_{f,k}}{\sigma_{f,k}} + \frac{\sum_{i=1}^{n_{f,k}}(G_{i,f,k}-\alpha-\gamma-\gamma_k-\eta_k)^2}{\sigma_{f,k}^3}, \\
        \frac{\partial\iota}{\partial\gamma_k} & = \sum_{k=1}^{K-1}\frac{\sum_{i=1}^{n_{m,k}}(G_{i,m,k}-\alpha-\gamma_k)}{\sigma_{m,k}^2} + \sum_{k=1}^{K-1}\frac{\sum_{i=1}^{n_{f,k}}(G_{i,f,k}-\alpha-\gamma-\gamma_k-\eta_k)}{\sigma_{f,k}^2}, \\
        \frac{\partial\iota}{\partial\eta_k} & = \sum_{k=1}^{K-1}\frac{\sum_{i=1}^{n_{f,k}}(G_{i,f,k}-\alpha-\gamma-\gamma_k-\eta_k)}{\sigma_{f,k}^2}, \\
        \frac{\partial\iota}{\partial\alpha} & = \sum_{k=1}^{K-1}\frac{\sum_{i=1}^{n_{m,k}}(G_{i,m,k}-\alpha-\gamma_k)}{\sigma_{m,k}^2} + \sum_{k=1}^{K-1}\frac{\sum_{i=1}^{n_{f,k}}(G_{i,f,k}-\alpha-\gamma-\gamma_k-\eta_k)}{\sigma_{f,k}^2} \\
         & + \frac{\sum_{i=1}^{n_{m,0}}(G_{i,m,0}-\alpha)}{\sigma_{m,0}^2} + \frac{\sum_{i=1}^{n_{f,0}}(G_{i,f,0}-\alpha-\gamma)}{\sigma_{f,0}^2}, \\
        \frac{\partial\iota}{\partial\gamma} & = \sum_{k=1}^{K-1}\frac{\sum_{i=1}^{n_{f,k}}(G_{i,f,k}-\alpha-\gamma-\gamma_k-\eta_k)}{\sigma_{f,k}^2} + \frac{\sum_{i=1}^{n_{f,0}}(G_{i,f,0}-\alpha-\gamma)}{\sigma_{f,0}^2},
    \end{aligned}
\end{equation}

Similarly, the Fisher information matrix can be well separated into two parts, the one for $(\sigma_{m,k},\sigma_{f,k})$ and the one for $(\alpha,\gamma,\gamma_k,\eta_k)$. Since the test does not involve $(\sigma_{m,k},\sigma_{f,k})$, we only require the second part of the Fisher matrix, which is of size $2K$ by $2K$.

Let the parameter set be $\theta=(\alpha,\gamma,\gamma_1,\eta_1,\dots,\gamma_{K-1},\eta_{K-1})'$. Before we introduce the Fisher information matrix, define quantities $t_k=\frac{n_{f,k}}{\sigma_{f,k}^2}+\frac{n_{m,k}}{\sigma_{m,k}^2}$ and $c_k=\frac{n_{f,k}}{\sigma_{f,k}^2}$.

The Fisher information matrix $I_n(\theta)$ is given by
\begin{equation*}
I_n(\theta) = \left(\begin{array}{@{}cc|cc|cc|cc@{}}
                \sum_{s=0}^r t_s & \sum_{s=0}^r c_s & t_1 & c_1 & t_2 & c_2 & \cdots & \cdots \\
                \sum_{s=0}^r c_s & \sum_{s=0}^r c_s & c_1 & c_1 & c_2 & c_2 & \cdots & \cdots \\\hline
                t_1 & c_1 & t_1 & c_1 &  &  &  &  \\
                c_1 & c_1 & c_1 & c_1 &  &  &  &  \\\hline
                t_2 & c_2 &     &     & t_2 & c_2 &  & \\
                c_2 & c_2 &     &     & c_2 & c_2 &  & \\ \hline
                \vdots & \vdots & & & & & \ddots &       \\
                \vdots & \vdots & & & & & & \ddots      
              \end{array}\right)
\end{equation*}

We can easily tell that $I_n(\theta)$ can be transformed into a block diagonal matrix through proper row and column transformations, and vice versa. This property is particularly helpful when we need to take the inverse of $I_n(\theta)$ for the Wald test.

Specifically, we can write
$$I_n(\theta) = R J_n(\theta) R',$$
where
\begin{equation*}
J_n(\theta) = \left(\begin{array}{@{}cc|cc|cc|cc@{}}
                t_0 & c_0 &  &  &  &  &  &  \\
                c_0 & c_0 &  &  &  &  &  &  \\\hline
                 &  & t_1 & c_1 &  &  &  &  \\
                 &  & c_1 & c_1 &  &  &  &  \\\hline
                 &  &     &     & t_2 & c_2 &  & \\
                 &  &     &     & c_2 & c_2 &  & \\ \hline
                 &  & & & & & \ddots &       \\
                 &  & & & & & & \ddots      
              \end{array}\right)
\end{equation*}
and 
\begin{equation*}
R = \left(\begin{array}{cccccc}
                1 &  & 1 & 0 & 1 & \cdots  \\
                 & 1 & 0 & 1 & 0 & \cdots  \\
                 &  & 1 &  &  &   \\
                 &  &  & 1 &  &   \\
                 &  &  &  & 1 & \\
                 &  &  &  &  & \ddots  
              \end{array}\right),
R^{-1} = \left(\begin{array}{cccccc}
                1 &  & -1 & 0 & -1 & \cdots  \\
                 & 1 & 0 & -1 & 0 & \cdots  \\
                 &  & 1 &  &  &   \\
                 &  &  & 1 &  &   \\
                 &  &  &  & 1 & \\
                 &  &  &  &  & \ddots  
              \end{array}\right).
\end{equation*}
The inverse of $R'$ is the transpose of $R^{-1}$.

Applying matrix $R$ to the left of any matrix is equivalent to adding the rows of odd indexes to the first row, and adding the rows of even indexes to the second row. Applying $R'$ to the right has the same effect but for the columns.

The inverse of the Fisher information matrix goes as
$$I_n^{-1}(\theta) = R'^{-1} J_n^{-1}(\theta) R^{-1}.$$
In this way, we successfully transform the problem of inverting a complicated matrix to the problem of inverting a block diagonal matrix.

After obtaining $J_n^{-1}(\theta)$, we perform row and column transformations to $J_n^{-1}(\theta)$. Applying matrix $R^{-1}$ to the right of any matrix is equivalent to extracting the first column from the columns of odd indexes, and extracting the second column from the columns of even indexes. Applying $R'^{-1}$ to the left has the same effect but for the rows.

Denote the individual blocks in $J_n(\theta)$ as
\begin{equation*}
    B_k = \left(\begin{array}{cc}
        t_k=\frac{n_{f,k}}{\sigma_{f,k}^2}+\frac{n_{m,k}}{\sigma_{m,k}^2} & c_k=\frac{n_{f,k}}{\sigma_{f,k}^2} \\
        c_k=\frac{n_{f,k}}{\sigma_{f,k}^2} & c_k=\frac{n_{f,k}}{\sigma_{f,k}^2}
    \end{array}\right).
\end{equation*}
We can further obtain
\begin{equation*}
    B_s^{-1} = \left(\begin{array}{cc}
        \frac{\sigma_{m,k}^2}{n_{m,k}} & -\frac{\sigma_{m,k}^2}{n_{m,k}} \\
        -\frac{\sigma_{m,k}^2}{n_{m,k}} & \frac{\sigma_{m,k}^2}{n_{m,k}} + \frac{\sigma_{f,k}^2}{n_{f,k}}
    \end{array}\right).
\end{equation*}

Sequentially, we can write $I_n^{-1}(\theta)$ as
\begin{equation}\label{fi}
I_n^{-1}(\theta) = \left(\begin{array}{cccc}
                 B_0^{-1} & -B_0^{-1} & -B_0^{-1} &  \cdots  \\
                 -B_0^{-1} &  B_1^{-1}+B_0^{-1} & B_0^{-1} &  \cdots  \\
                 -B_0^{-1} &  B_0^{-1} & B_2^{-1}+B_0^{-1} &  \cdots  \\
                 \vdots & \vdots &  \vdots   & \ddots   
              \end{array}\right)    
\end{equation}

With Eq \eqref{fi} in hand, we can construct the Wald statistics for any combination of parameters from the set $(\alpha,\gamma,\gamma_1,\eta_1,\dots,\gamma_{K-1},\eta_{K-1})'$.

\section{Equivalency of testing $H_0:\eta_l=0$ and $H_0:\eta_k-\eta_l=0$ by redefining the baseline population}

Since the inverse of the Fisher information matrix is already given in Eq \eqref{fi}, by plugging in the corresponding quantities into the expression of a Wald statistic, we can easily show that both testing $H_0:\eta_l=0$ and testing $H_0:\eta_k-\eta_l=0$ yield the test statistics (2.16) and (2.17) from the manuscript, thus they are equivalent.